\documentclass[a4paper,twoside,12pt]{article}

\usepackage{graphicx}

\def\epsi{\epsilon}

\def\A{{\cal A}}

\def\R{{\cal R}}

\def\M{{\cal M}}
\def\li2{{\rm Li}_2}

\def\vec{\overrightarrow}

\def\roughly#1{\,\,\raise.3ex\hbox{$#1$\kern-.75em\lower1ex\hbox{$\sim$}}\,\,}

\def \lsim{\mathrel{\vcenter
     {\hbox{$<$}\nointerlineskip\hbox{$\sim$}}}}
\def \gsim{\mathrel{\vcenter
     {\hbox{$>$}\nointerlineskip\hbox{$\sim$}}}}

\def\fo{\hbox{{1}\kern-.25em\hbox{l}}}

\def\bea{\begin{eqnarray}}
\def\eea{\end{eqnarray}}
\def\beq{\begin{equation}}
\def\eeq{\end{equation}}
\def\eq{\end{equation}}
\def\to{\rightarrow}

\def\bsg{\ifmmode B\to X_s\gamma\else $B\to X_s\gamma$\fi}
\def\bsll{\ifmmode B\to X_s\ell^+\ell^-\else $B\to X_s\ell^+\ell^-$\fi}
\def\bstt{\ifmmode B\to X_s\tau^+\tau^-\else $B\to X_s\tau^+\tau^-$\fi}
\def\shat{\ifmmode \hat{s}\else $\hat{s}$\fi}

\def\Emiss{\not  \! \! E}
\def\jet{{\rm jet}}

\newcommand{\newc}{\newcommand}

\newc{\lcal}{\int {\cal L}dt}
 
\newc{\LSP}{{\chi^0_1}}
\newc{\stauR}{{\tilde \tau_R}}
\newc{\stau}{{\tilde \tau_1}}
\newc{\mstop}{m_{\tilde{t}}}
\newc{\mHpm}{m_{H^\pm}}
\newc{\ie}{{\it i.e.}}          
\newc{\etal}{{\it et al.}}
\newc{\eg}{{\it e.g.}}          
\newc{\kev}{\hbox{\rm\,keV}}            
\newc{\mev}{\hbox{\rm\,MeV}}            
\newc{\gev}{\hbox{\rm\,GeV}}            
\newc{\tev}{\hbox{\rm\,TeV}}
\newc{\xpb}{\hbox{\rm\, pb}}
\newc{\xfb}{\hbox{\rm\, fb}}

\newc{\mtop}{m_t}
\newc{\mbot}{m_b}
\newc{\mz}{m_Z}
\newc{\mw}{M_W}
\newc{\alphasmz}{\alpha_s(m_Z^2)}
\newc{\swsq}{\sin^2\theta_W}
\newc{\tw}{\tan\theta_W}
\newc{\cw}{\cos\theta_W}
\newc{\sw}{\sin\theta_W}
\newc{\BR}{\hbox{\rm BR}}
\newc{\zbb}{Z\to b\bar}
\newc{\Gb}{\Gamma (Z\to b\bar b)}
\newc{\Gh}{\Gamma (Z\to \hbox{\rm hadrons})}
\newc{\rbsm}{R_b^\hbox{\rm sm}}
\newc{\rbsusy}{R_b^\hbox{\rm susy}}
\newc{\drb}{\delta R_b}

\newc{\sgn}{\mbox{sgn}}
\newc{\tbeta}{\tan\beta}
\newc{\uL}{{\tilde u_L}}
\newc{\uR}{{\tilde u_R}}
\newc{\cL}{{\tilde c_L}}
\newc{\cR}{{\tilde c_R}}
\newc{\tL}{{\tilde t_L}}
\newc{\tR}{{\tilde t_R}}
\newc{\dL}{{\tilde d_L}}
\newc{\dR}{{\tilde d_R}}
\newc{\sL}{{\tilde s_L}}
\newc{\sR}{{\tilde s_R}}
\newc{\bL}{{\tilde b_L}}
\newc{\bR}{{\tilde b_R}}
\newc{\eL}{{\tilde e_L}}
\newc{\eR}{{\tilde e_R}}
\newc{\mhp}{m_{H^\pm}}
\newc{\mhalf}{m_{1/2}}
\newc{\emt}{{e/\mu /\tau}}
\newc{\MGRW}{{M_D^{\scriptscriptstyle GRW}}}

\newc{\bW}{{\bar W}}
\newc{\bB}{{\bar B}}
\newc{\eps}{{\epsilon}}
\newc{\bg}{{\bar g}}

\def\lappeq{\mathrel{\rlap{\raise.5ex\hbox{$<$}}
{\lower.5ex\hbox{$\sim$}}}}
\def\gappeq{\mathrel{\rlap {\raise.5ex\hbox{$>$}}
{\lower.5ex\hbox{$\sim$}}}}

\newcommand{\drawsquare}[2]{\hbox{%
\rule{#2pt}{#1pt}\hskip-#2pt
\rule{#1pt}{#2pt}\hskip-#1pt
\rule[#1pt]{#1pt}{#2pt}}\rule[#1pt]{#2pt}{#2pt}\hskip-#2pt
\rule{#2pt}{#1pt}}

\newcommand{\Dal}{\drawsquare{7}{0.6}}

\newc{\etamn}{\eta_{\mu\nu}} 
\newc{\etars}{\eta_{\rho\sigma}}
\newc{\etamr}{\eta_{\mu\rho}}
\newc{\etarn}{\eta_{\rho\nu}}
\newc{\etams}{\eta_{\mu\sigma}}
\newc{\etasn}{\eta_{\sigma\nu}}

\vspace{20pt}

\begin{document}

\begin{flushright}
CERN-PH-TH/2006-029\\
hep-ph/0607055
\end{flushright}

\hfill

\vspace{20pt}

\begin{center}
{\Large \textbf
{Cargese Lectures
on  Extra Dimensions}}
\end{center}

\vspace{6pt}

\begin{center}
\textsl{ 
 Riccardo Rattazzi} \vspace{20pt}

\textit{Department of Physics, CERN, 
 CH-1211 Geneva 23, Switzerland}
\end{center}

\vspace{12pt}

\begin{center}
\textbf{Abstract }
\end{center}

\vspace{4pt} {\small \noindent 
I give a pedagogical introduction to the concepts and the tools
that are necessary to study particle physics models in higher
dimensions. I then give a more detailed presentation of warped
compactifications and discuss their possible relevance to the hierarchy
problem.}
\vskip 2.0truecm
{\it Lectures at the Cargese School of Physics and Cosmology, August 4th -16th 2003}
\vfill\eject 
\noindent

\section*{Prologue}
Science in general and particle physics in particular thrive from
conceptual puzzles and unexplained phenomena. The gauge hierarchy problem
is an exemplar source for inspiration. While we haven't got yet any direct experimental evidence onto what
mechanism sets the  small ratio $G_N/G_F\sim 10^{-34}$ between the Newton and Fermi constants,
 a great deal of theoretical progress in particle physics has been
triggered in trying to come up with an explanation. For instance, the great development
in supersymmetric field theory of the last three decades is to a good extent
motivated by the potential relevance of supersymmetry to the hierarchy
problem. The last few years have also witnessed a great revival in the interest for 
models with extra space dimensions. On one side this revival is motivated by important
theoretical developments within superstring theory, in particular by the realization that
there exist in string theory solitonic membranes, D-branes, on which ordinary
particles could be localized \cite{Polchinski:1996na}. On the other side the revival is also 
phenomenologically motivated by
the realization that extra dimensions can shed a new light on the hierarchy problem \cite{add}.

The potential relevance of extra dimensions to the hierarchy problem can be
grasped by the following simple line of reasoning. One way to phrase the hierarchy is
that the Standard Model (SM) quanta, like the $Z$-boson, are much softer than the quanta of a
possible underlying Grand Unified Theory (GUT) or string theory: $m_Z/m_{GUT}\sim 10^{-14}$.
That is to say that the minimum frequency corresponding to a travelling $Z$ boson wave is $m_Z\sim 10^2$ GeV
while the minimal frequency of a GUT wave is $\sim 10^{16}$ GeV. Supersymmetry or technicolor allow for a
dynamical explanation of this huge hierarchy. Moreover in both cases the value of $m_Z/m_{GUT}$ is determined
by a quantum phenomenon, {\it i.e.} dimensional transmutation. However we know since long of a basic classical
phenomenon that can make quanta softer: gravitational Redshift. Let us briefly recall how this works
in general relativity. Consider a gravitational field specified by a metric $g_{\mu\nu}(x)$. The invariant interval
separating event $x$ from event $x+dx$ is given by $(ds)^2=g_{\mu\nu}dx^\mu dx^\nu$. By the Equivalence Principle
$(ds)^2$ equals the Lorentz invariant interval measured by any freely falling
observer at $x$: $(ds)^2=-(\Delta X_0)^2+(\Delta X_1)^2+(\Delta X_2)^2+(\Delta X_3)^2$. Consider the case of a static
metric $g_{\mu\nu}$ and of an interval $dx$ in the time direction. In this case $(ds)^2=g_{00}(dx^0)^2< 0$ corresponds to the proper
time interval experienced by a freely falling observer with zero velocity at $x$
\beq
d\tau\,=\,\sqrt {-g_{00}(x)} dx^0
\label{delay}
\eeq
By the Equivalence Principle any clock at rest at $x$ will oscillate with its proper period $\Delta T=1/\omega$
according to the same freely falling observer. Then by eq. (\ref{delay}) the frequency $\omega(x)$ oberved in the 
original reference frame will be rescaled according to
\beq
\frac{1}{\omega}\,=\,\sqrt {-g_{00}(x)} \frac{1}{\omega(x)}.
\label{delay2}
\eeq
\begin{figure}[t]
\vskip 20pt
\begin{center} 
$$\includegraphics[width=10.5cm]
{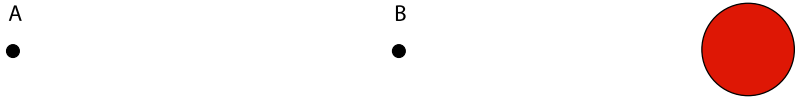}$$
\caption{\em Two atoms A and B in the gravitational field of a star.}
\label{star2}
\end{center}
\end{figure}
This rescaling is not yet, by itself, a physical effect: also the unit of measure of time at $x$,
specified by some standard clock, will undergo the same rescaling. (Another way to state this is that
the overall normalization of the metric, upon which eq.~(\ref{delay2}) crucially depends, is not an observable
as it depends on the units of measure.) An observable effect arises when comparing the frequencies of
two  copies of the same clock  located at different points. For instance we can consider two
hydrogen atoms located respectively at points A and B in the gravitational field of a star (see Fig.~(\ref{star2})),
and associate the frequency $\omega$ to a given atomic transition.
The waves emitted at A will oscillate everywhere with frequency $ \omega_A=\sqrt{-g_{00}(A)} \omega $.
This is because the background is time independent, and two neighbouring wavecrests, leaving A with
delay $1/\omega_A$, will take the same time to reach any given point, for example point B.
Similarly the wave emitted at B will oscillate everywhere with frequency $ \omega_B=\sqrt{-g_{00}(B)} \omega $.
The ratio
\beq
\frac{\omega_B}{\omega_A}=\sqrt{\frac{g_{00}(B)}{g_{00}(A)}}<1
\eeq
represents a physical effect. The observer A notices that the light emitted from position B, closer to the star,
is redshifted and similarly observer B notices that the light emitted at A is blue-shifted. This effect can be 
qualitatively
understood as the  photon loosing kinetic energy as it climbs up to A from deep inside the gravitational potential 
well
at B. In everyday's life this gravitational redshift represents a tiny effect, since the gravitational field of the earth
is rather weak. However gravity is a non-linear theory encompassing large gravitational fields, like the one
near the horizon of a black hole. Assume indeed that the star of the previous example has collapsed to
form a black hole. In this case the metric is given as a function of the radial coordinate $r$ by 
$g_{00}=-1+2G_N M/r\equiv -1+r_S/r$ (with $M$ the black hole mass). As the position $r_B$ of atom B approaches the 
horizon $r_S$ we have
$g_{00}(r_B)\to 0$ so that the the Redshift becomes infinite! The infinite Redshift of
 photons emitted at the horizon corresponds to the fact  that light cannot escape from a black-hole. We can say that
the gravitational field of a black hole creates an infinite {\it hierarchy} of energies of the emitted photons
as the emitter is moved towards the horizon. With the gauge hierarchy in  mind it is perhaps then natural to think 
of a wild generalization
of the system we just discussed, one in which the points A or B are generalized to 3-dimensional spacelike surfaces, or 3-branes.
In this process the dynamical system living on a point, the atom, is generalized to the dynamical system that
lives on a 3-surface, a 3+1 quantum field theory, for instance the Standard Model. Imagine then to place two 
identical copies
of 3-branes hosting the Standard Model at different points inside a gravitational field, 
 in Fig.~(\ref{horizon}), in
a straightforward generalization of Fig.~(\ref{star2}). Of course, since three space dimensions already span the 
membranes,
the distance separating them must correspond to a new spacelike dimension, the fifth dimension. In our 
generalization
the role of the atomic energy levels (and thus of the emitted frequencies) is played by the masses of the particles
living on the branes. For instance the masses of the two identical $Z$ bosons  satisfy
\beq
\frac{m_Z(B)}{m_Z(A)}=\sqrt{\frac{g_{00}(B)}{g_{00}(A)}}<1.
\label{redz}
\eeq
\begin{figure}
\begin{center}
\includegraphics[width=9cm]{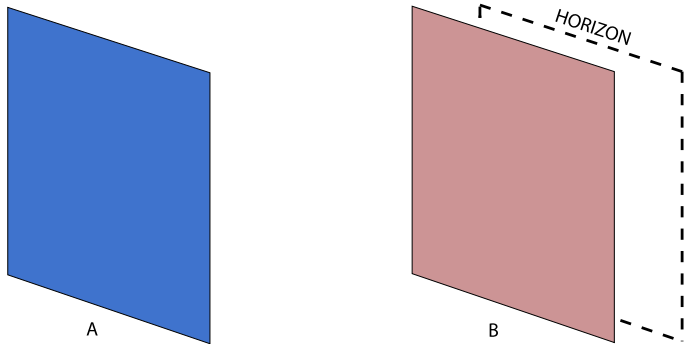}
\end{center}
\caption{Two branes A and B in a 5-dimensional gravitational field.\label{horizon}}
\end{figure}
If only gravity propagates in the fifth dimension, one experimental consequence of eq.~(\ref{redz}) 
is that the gravitons emitted in radiative $Z$ decays at point A are  more energetic to observer B
than those emitted in the same process at B. Also one can't stop from imagining 
a situation where brane B is much deeper than brane A inside a gravitational field, perhaps even very
close to a horizon: then one would expect a huge hierarchy for the masses of identical particles living
on the two different branes. 
Of course the example we are considering is not directly applicable to the gauge hierachy problem, as that does not 
concern
two identical copies of the SM. Nevertheless the Redshifting mechanism would obviously be at  work even
if the field theories living on the two branes where not the same, and also in more complicated situations
where  the SM degrees of freedom are not fully localized: it is just kinematics.
Is it then possible that the weak scale hierarchy originates as a consequence of gravitational Redshift
in extra-dimensions? The answer to this question is affirmative and the model that proves it was proposed
by Randall and Sundrum in a pioneering paper \cite{rs}.

\section{Part I: Extra Dimensions}
In this section we introduce the players in the game (gravity, branes and localized fields)
and discuss the rules that govern their effective action.

\subsection{Gravity and Branes}
\label{Gravity and Branes}
Gravity plays a central role in the physics of extra dimensions. This is shown for instance by the example
of the previous section. It is then important to recall the basic concepts relavant to describing
the dynamics of gravity. A basic introduction to General Relativity is taken for granted here.
We will be concerned with a D-dimensional space-time with coordinates $X^{M}$ and metric $g_{MN}(X)$, with $M,N=0,\dots, D-1$.
We use the mostly plus convention wherein at any space time point   a locally
inertial frame can be found in which $g_{MN}=\eta_{MN}\equiv(-1,+1,\dots,+1)$. Later on we will specialize
to the phenomenologically relevant case where $D-4$ space-like dimensions are compactified.
The metric, up to diffeomorphisms, contains the dynamical degrees of freedom of gravity.
The affine connection $\Gamma_{MN}^R=g^{RS}(\partial_M g_{NS}+\partial_N g_{MS}-\partial_Sg_{MN})/2$
defines parallel transport, by means of which the  Riemann tensor, characterizing the spacetime curvature, is constructed
$\R_{MNR}^S=\partial_N\Gamma_{MR}^S-\partial_M\Gamma^S_{NR}+\Gamma^T_{MR}\Gamma^S_{TN}-\Gamma^T_{NR}\Gamma^S_{TM}$.
Using the Riemann tensor and its contracted forms, the Ricci tensor $\R_{MR}=\R_{MNR}^{N}$ and Ricci scalar $\R=\R_{MN}g^{MN}$,
the most general invariant action can be written as
\beq
S=\int d^DX \sqrt{g}\left \{ aM_D^D+2M_D^{D-2} \R+b M_D^{D-4}\R^2+\dots cM_D^{D-6} \R\Dal \R+\dots\right \}
\label{Lgravity}
\eeq
where $M_D$ is $D$-dimensional Planck scale. In ordinary 4-dimensional Einstein gravity, according to our 
non-conventional normalization, we have $M_4= (32 \pi G_N)^{-1/2}\simeq 1.2\times 10^{18}$ GeV.
We  have parametrized all the couplings with dimensionless
coefficients $a,b,c, \dots$ and organized the lagrangian as a derivative expansion
\beq
aM_D^D p^0+ M_D^{D-2}p^2 +b M_D^{D-4} p^4+\dots+c M_D^{D-6}p^6+\dots
\eeq
where the lowest terms are those that are important at the longest distance scales.
 In particular the first term, the cosmological
constant, influences directly the global structure of the space-time. 
From the effective lagrangian point of view, which we will discuss in more detail later, we can  meaningfully
address only those phenomena that involve a finite number of terms, i.e. those for which $p/M_D$ is significatively less than 1.
 In this respect, although one would
naively expect all the coefficients $a,b,c, \dots$ to be $O(1)$, we will assume 
that the cosmological term $a$  is much smaller than 1. This is mostly for a theoretical reason: when $a=O(1)$ the background 
solution to Einstein equations
has $\R=O(M_D^2)$ for which the derivative expansion breaks down. But notice also that there is a 
phenomenological preference to work with small cosmological constant. On one side we know from observation that the effective
cosmological constant of our macroscopic 4-dimensional world is very small: 
$\Lambda^4=a_4M_4^4\sim (10^{-3} {\rm eV})^4\ll M_4^4$.
This  may be an indication that also the fundamental cosmological constant $a M_D^{D-2}$, before compactification,
is small. Moreover, as mentioned in sec.~\ref{LED}, a small $D$-dimensional cosmological constant 
is also favored  in the scenario of large extra dimensions \cite{add} by the requirement of a flat potential for the 
radius modulus. So, while we will assume $a\ll 1$ as a result of some tuning or, perhaps, 
$D$-dimensional supersymmetry,
for all the other coefficients we will just need the perfectly natural and weak assumption that they be $\leq O(1)$.
For instance in ordinary General Relativity with the above assumption the Einstein Lagrangian
is a successful truncation up to the very small Planck length  $\lambda_P =1/M_4=10^{-33}$ cm. This is evident with the above
classical Lagrangian, but as discussed later, it remains true also at the quantum level.
\vskip0.3truecm
The second important player in the extra dimensional game is given by the so called (mem)branes.
They are extended objects which  span surfaces and on which excitations (particles) can 
be localized. An explicit physical example of a brane is given, for instance, by the surface separating two different metals,
where there exist localized excitations in the charge density, the surface plasmons.
 Another explicit example
of a brane can be provided by a domain wall. Consider a scalar field theory with a $Z_2$ invariant potential
$V(\phi)=\frac{g}{2}(\phi^2-v^2)^2$. In addition to the two vacuum solutions $\phi = \pm v$ this model
contains domain wall solutions
\beq
\phi = v \tanh(m (z-z_0))
\label{wall}
\eeq
where $m={\sqrt g} v$ is the mass of the scalar field, while $z$  indicates one of the space directions. 
This solution interpolates between the $\phi=-v$ vacuum
at $z=-\infty$ and the $\phi = +v$ vacumm at $z=+\infty$ and is thus topologically stable \cite{rajaraman}.
With respect to the true vacuum $\phi = \pm v$ this solution has an energy density $E= (\partial_z \phi)^2\propto
\cosh^{-4}(m(z-z_0)$ localized within a distance $\sim 1/m$ from the center of the wall. Integrating $E$ across the wall
we obtain the wall tension $\tau = 4 {\sqrt g} v^3/3$.
Away from the wall the scalar has mass $m$ so that at $E<m$ there are no modes propagating through the full space.
There is however a massless scalar mode localized at the wall. Indeed the domain wall solution breaks
spontaneously the original $D$-dimensional Poincar\'e group down to the $(D-1)$-dimensional one, corresponding
to translations and boosts in the directions parallel to the wall. As it happens for ordinary internal symmetries,
we then expect the presence of Goldstone bosons associated to the broken generators.
 A naive application of that result to spacetime symmetries is however not possible\cite{Sundrum:1998sj,Low:2001bw}. 
The Goldstone theorem is proven by considering
the local tranformations associated to the global symmetries. In the case of the Poincar\'e
group both translations and boosts reduce, locally, to local translations. The Goldstone bosons
are then in a one to one correspondence with the broken translation generators \cite{Sundrum:1998sj}. 
In the case at hand the translations along the
$z$ direction are broken and $z_0$ parametrizes the manifold of equivalent vacua. Like in the case of internal symmetries
we can parametrize the Goldstone excitation by promoting $z_0$ to a field $z_0(x)$ depending on the $(D-1)$ longitudinal coordinates.
At linearized level it corresponds to a  mode 
\beq
\delta \phi(x,z)=-\frac{vm}{\cosh^2 m z} \,z_0(x)
\eeq
which is clearly normalizable and localized within a distance $1/m$ around the wall.  Under $z$-translations $\phi(z,x)\to \phi(z+a,x)$
we have $z_0(x)\to z_0(x)-a$. Because of this non linear symmetry the action can depend on $z_0$ only through its space
time derivatives. In particular there is no mass term for $z_0(x)$. The effective action for $z_0$, valid at momentum $\ll m$, can be
carefully derived by integrating out the massive excitations of the original field $\phi$. However it is intuitively clear what
result to expect at lowest order. In this limit
we are considering very smooth deformations of the wall,  such that its position varies appreciably only
over distances  much bigger that its width $1/m$.  
At each point we then expect the field to be given approximately by eq.~(\ref{wall}), but with $z$ replaced 
by the direction locally orthogonal to the wall. Then the action will just be given by the integral 
of the original wall tension $\tau={\sqrt \lambda} v^3$ over the volume of the deformed wall. 

Other fields can be localized at a domain wall. A fermion $\psi$ with Yukawa interaction $\lambda \psi \bar \psi$ wil be massive
in the bulk, but will contain zero modes localized at the wall where $\phi=0$ \cite{Rubakov:bb}. Similarly, 
ideas to localized gauge fields have been
proposed \cite{Dvali:1996xe}. Likewise, D-branes in string theory support localized modes (scalars, fermions and vectors) 
associated to open strings ending on them \cite{Polchinski:1996na}.
In these lectures we will be focusing on the low energy description of branes. For instance, 
in the case we just considered
this corresponds to $E\ll m$. In this regime we will
not be concerned with the microscopic mechanism that gave origin to the brane and to the fields localized on it. We will
just assume that the brane hosts a field theory of our choice and derive the consequences. 
The presence of some degrees of freedom, like the Goldstone $z_0$
above, could however just follow from symmetry considerations and not be an option.

\subsection{Brane Effective Actions}

We will now discuss the dynamics of branes by writing the most general effective action
satisfying some basic principles. We follow closely the presentation given in ref.~\cite{Sundrum:1998sj}. 
Let us consider an n-brane, a membrane filling $n$ spacial
dimension whose spacetime trajectory, the worldbrane, is an $n+1$-surface. We parametrize this surface with
coordinates $x^\mu$, $\mu=0,\dots,n$. The embedding in the full $D$-dimensional spacetime
is described by $D$ functions $X(x)^M$, with $M=0,\dots,D-1$. For instance, in the simplest case of a point particle,
a 0-brane, the worldbrane is  the particle trajectory, the worldline, parametrized by a time coordinate $x^0$: $X(x^0)$.
It is physically intuitive that the distance between points on the brane, as measured by a brane observer,
be the same as measured by a bulk observer, 
\bea
ds^2|_{brane}&=&G_{MN}\left ( X(x)\right ) dX|_{brane} dX^N|_{brane}\cr
&=&G_{MN}\left ( X(x)\right )\partial_\mu X\partial_\nu X^N dx^\mu dx^\nu
\eea
{\it i.e.} the bulk metric gives rise to an induced metric ${\hat g}_{\mu\nu}$ on the brane
\beq
{\hat g}_{\mu\nu}(x)=G_{MN}\left ( X(x)\right )\partial_\mu X\partial_\nu X^N.
\label{induced}
\eeq
Notice that the induced metric is a scalar under the bulk diffeomorphisms
(all the $M,N...$ indices are contracted) while it is a tensor under reparametrizations of the brane $x^\mu= {x}^\mu(x')$.
As the choice of coordinates $x$ is arbitrary, physical quantities should not depend on it. 
Therefore, like in ordinary gravity, starting from ${\hat g}_{\mu\nu}$ we should
write an action invariant under brane reparametrizations. We will do that in a moment. Before then we want
to emphasize that,
like we have projected the metric, so we can do with other tensors. For instance   
a bulk gauge field (1-form) $A_M(X)$ leads to a brane field
${\hat A}_\mu(x)=A_M\left ( X(x)\right )\partial_\mu X^M$ which under the bulk gauge tranformation $A_M \to A_M +\partial_M \alpha$
shifts as under a proper $n+1$-dimensional gauge transformation
\beq
\delta{\hat A}_\mu(x)=\partial_M\alpha\left (X(x)\right )\partial_\mu X^M(x)=\partial_\mu\left [\alpha
\left (X(x)\right )\right ].
\eeq
We can then use the projected field and gauge symmetry to couple the original 1-form to charged matter on the brane.
A similar procedure can be followed for the D-bein field $E^A_M$, necessary to couple fermions to gravity
in a manifestly covariant way. Here and in what follows we indicate with  $A,B,\dots$, $A=0,\dots D-1$, and with
$a,b\dots$, $a=0,n$ respectively the bulk and brane Lorentz indices. 
The D-bein $E^A_M$ field represents D  1-forms ($A=0,\dots,D-1$) in the cotangent space, satisfying the relation
$\eta_{AB}E^A_ME^B_N=G_{MN}$.  
The D-bein defines at each space-time point a tangent space basis corresponding to the coordinates of a free falling observer: it
 associates to the entries of a vector $V^M$ in  a given system of coordinates
the entries $\hat V^A=E^A_M V^M$ in the coordinates of a free falling observer.
Indeed, by the definition of $E^A_M$, vector products are conserved: $\hat V^A\hat W^B \eta_{AB}=V^MW^NG_{MN}$. Moreover local Lorentz rotations 
$E^A_M(X) \to R^A_B(X)E^B_M(X)$ are a gauge symmetry: the 
orientation of the $D$-bein at each point is not physical. This just means that the locally inertial reference frame is only defined up
to a Lorentz transformation.
Now, the tangent space $\sigma$ to the brane at a point $x$ is a $n+1$ subspace of the tangent space $\Sigma$ at $X(x)$.
A vector $v^\mu\in {\sigma}$ is written in free falling coordinates as $\hat v^A=E^A_M\partial_\mu X^M v^\mu$. By this relation, $\sigma$
is represented as a $n+1$ dimensional  subspace of the Lorentzian (free falling) vector space.
To define the induced $n+1$-bein we have just to find an orthonormal basis of this subspace.
One way to proceed is to
 divide the indices $\{ A\}$ into two groups: $\{ a\} $ for $A=0,\dots,n$ and $\{i\}$ for $A=n+1,\dots,D-1$.
Since $\sigma$ is a time-like subspace we can always perform a Lorentz rotation $\hat {v}'^A= \bar R^A_B \hat v^B$
such that $\hat{v}'^i\equiv 0$ for ${\hat v}^A\in \sigma$. In the new basis, $\sigma$ is spanned by ${\hat v}'^a $ for  $a=0,\dots, n$, so that we have
$\hat{v}'^a\hat {v}'^b\eta_{ab}\equiv \hat{v}'^A\hat {v}'^B\eta_{AB}=v^\mu v^\nu g_{\mu\nu}$. For vectors in $\sigma$,
summing over $a$ is equivalent to  summing over $A$.
The induced $n+1$-bein can  then be defined as $e^{a}_\mu \equiv \bar R^a_B E^B_M\partial_\mu X^M$.
It is straightforward to check that $e^a_\mu$ satisfies the basic conditions
\beq
e^a_\mu e^b_\nu \eta_{ab}= {\hat g}_{\mu\nu}\quad\quad e^a_\mu e^b_\nu {\hat g}^{\mu\nu}=\eta^{ab}.
\eeq
 One crucial remark is that the rotation $\bar R^A_B$ is only
defined modulo the rotations in the subgroup $SO(1,n)\times SO(D-n-1)$ which leave the 
$\sigma$ subspace and its complement invariant. In particular the induced $e^a_\mu$ is defined modulo
the local Lorentz symmetry $SO(1,n)$ of the brane:  the arbitrariness of our construction of $e^a_\mu$
does not affect physics provided the brane Lagrangian is written in a locally Lorentz
invariant way. Using $e^a_\mu$ we can derive a brane spin connection and write a covariant Lagrangian for localized 
fermions.

The effective Lagrangian for a brane can then be written as
\bea
{\cal S}_{brane}=\int d^{n+1} x 
{\sqrt g}\Bigl\{ &-\tau + M^{n-1} \R(\hat g) +\bar \psi \not\! D \psi + D_\mu 
\phi^\dagger D^\mu \phi+ 
\cr 
&+\bigl (\hbox{{\rm higher derivatives}}\bigr )
\Bigr \}
\label{braneaction}
\eea
where we have considered the example of localized fermion and scalar fields. The covariant derivatives involve 
the projected 1-forms, ${\hat A}_\mu$ and $e^a_\mu$, as well as any possible localized Yang Mills field. The 
most relevant term,
the one with the lowest number of derivatives, corresponds to the brane tension $\tau$. In the effective theory 
description
of the domain wall of the previous section, $\tau$ is determined by matching eq.~(\ref{braneaction}) computed on a 
flat
configuration, on which the intrinsic curvature terms vanish, with our calculation of the flat wall tension. 
This way we obtain 
$\tau=4{\sqrt g}v^3/3$. The general result in eq.~(\ref{braneaction}) also shows that the naive derivation of the effective
action sketched in sect.~\ref{Gravity and Branes} is indeed accurate in the limit where the wall intrinsic curvature is small.

Let us consider as an explict example a 3-brane living in $D$ dimensional Minkowsky space, in the limit in which gravity
is turned off. We can choose a gauge where the brane embedding is simply $X^\mu=x^\mu$ for $\mu=0,1,2,3$
and $X^i=Y^i(x)$ for  $i=4,\dots,D-1$. In this parametrization the brane roughly
extends along the $0,\dots,3$ direction of the bulk space. The functions $Y^i$ parametrize the deformations along the orthogonal
directions and are the dynamical degrees of freedom, the branons.  In terms of the branons the induced metric is
\beq
{\hat g}_{\mu\nu}=\eta_{\mu\nu}+\partial_\mu Y^i\partial_\nu Y^j \delta_{ij}.
\label{Ymetric}
\eeq
and the tension term of eq. (\ref{braneaction}) expanded in powers of $\partial Y$ becomes
\beq
{\cal L}_{eff}= -\tau{\sqrt -g}=-\tau\left \{1+\frac{1}{2}\partial_\mu Y^i\partial^\mu Y_i+\frac{1}{8}( \partial_\mu Y^i\partial^\mu Y_i)^2+\dots\right \}.
\label{tensionexp}
\eeq
This lagrangian provides a kinetic term with the right sign for $Y$ provided $\tau>0$. The configuration $Y^i=0$ is stable and represents
the vacuum configuration of our brane. But the most remarkable thing is that our general symmetry considerations
also fix all the interaction terms involving $n$ fields $Y$ and a number of derivatives $\leq n$. The terms involving the curvature
affect the interactions that have always at least two more derivative, and give subleading contributions to the scattering
of branons at low enough energy. This is totally analogous to what happens in ordinary sigma models, like the pion lagrangian of QCD, where at lowest
order in the expansion $E/f_\pi$ the scattering amplitudes are fully fixed by the group structure
 in terms of just one physical parameter,
$f_\pi$ itself. Indeed also the branon system is a $\sigma$-model whose coset space corresponds to the breakdown of 
the $D$-dimensional Poincar\'e
group down to the $4$-dimensional one \cite{Sundrum:1998sj,Low:2001bw}. To conclude, notice that
by indicating $\tau=f^4$ and by going to the canonical field ${\hat Y^i}=f^2 Y^i$, the branon interactions are proportional to inverse
powers of $f$: this mass scale plays a role analogous to $f_\pi$ in the pion Lagrangian.

\subsection{Effective Field Theories}

One general aspect of physical systems is that the dynamics at large length scales, or
equivalently at low energy, does not depend too much on the microscopic details.
For instance the interaction of an electromagnetic wave with an antenna of size $a$ much smaller than the
 wavelength $\lambda$ is described to a good accuracy by the coupling to the dipole
mode of the antenna. Higher multipole moments  will contribute to corrections suppressed by
powers of $\lambda a\ll 1$. Another example is provided by molecules, where the {\it slow} vibrational modes,
describing the oscillations in the distance between the various nuclei, can
be accurately studied by first averaging over the {\it fast} motions of electrons.
Averaging over the electronic states provides an effective Hamiltonian for the low frequency
modes, where the higher details of the electronic structure are controlled by higher powers of the ratio $\omega_{slow}/\omega_{fast}$.
Effective Field Theories technique provide a systematic way, an expansion, to treat the details
of microscopic physics when discussing phenomena at low enough energy. Normally, when Quantum Field Theory is introduced as
a construction to describe fundamental processes, a great emphasis is put on the requirement of renormalizability.
Technically renormalizability corresponds to the possibility of sending the energy cut-off $\Lambda$  of the system 
to infinity while keeping all the physical quantities finite (and non trivial)\footnote{As we will explain better below this is a somewhat stronger requirement 
than renormalizability: in weakly coupled theories it corresponds to asymptotic freedom.}. Physically this means
that the theory can be extrapolated to infinitely small distances without encountering new microscopic structures. Renormalizable theories can be
truly {\it fundamental} and not just an {\it effective} description valid in a limited energy range.  
Renormalizable theories are however a special case, and in practically all applications to particle physics
one deals with non-renormalizable effective field theories \footnote{For  excellent introductions to effective field theories see the papers in Ref.\cite{effective}} 
The best example of a non-renormalizable theory is given by General Relativity, which necessarily requires a new 
description at an energy
scale smaller or equal to $M_P=10^{19}$ GeV. Nonetheless GR makes perfect sense as an effective field theory 
at energies much smaller than its cut off \cite{Donoghue:dn}. But also QED is an effective field theory: the interactions of electrons and photons
are modified at energies much bigger than $ m_e$ by the presence of new particles and new interactions. Still in the regime 
$E\sim m_e$
the {\it small} effects of the microscopic dynamics can be accounted for by adding a suitable tower of non-renormalizable interactions. 
 The SM, the most fundamental description of
particle interactions gravity excluded, is renormalizable. Still we can only consider the SM as an effective theory.
On one hand this is because the necessary inclusion of gravity makes it non-renormalizable. On the other hand,
even in the absence gravity, the SM is renormalizable but not asymptotically free. At least one of its couplings,
the one associated to the hypercharge vector boson, grows logarithmically with energy and becomes infinite at a scale
 $M_L\sim m_We ^{ b/\alpha} $, with $b= 12\pi \cos^2\theta_W/41\sim 1$. 
At $E\sim M_L$ the perturbative description breaks down, very much
like the effective description of G.R. breaks down at the Planck scale. The fact that $M_L\gg M_P$ makes however
this second problem academic.

\begin{figure}[t]
$$\includegraphics[width=8cm]{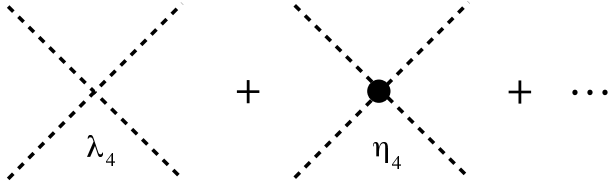}$$
\caption{\label{fourlegs}\em The diagrams contributing to the elastic process $2\to 2$ at lowest order.}
\end{figure}
\begin{figure}[t]
$$\includegraphics[width=12cm]{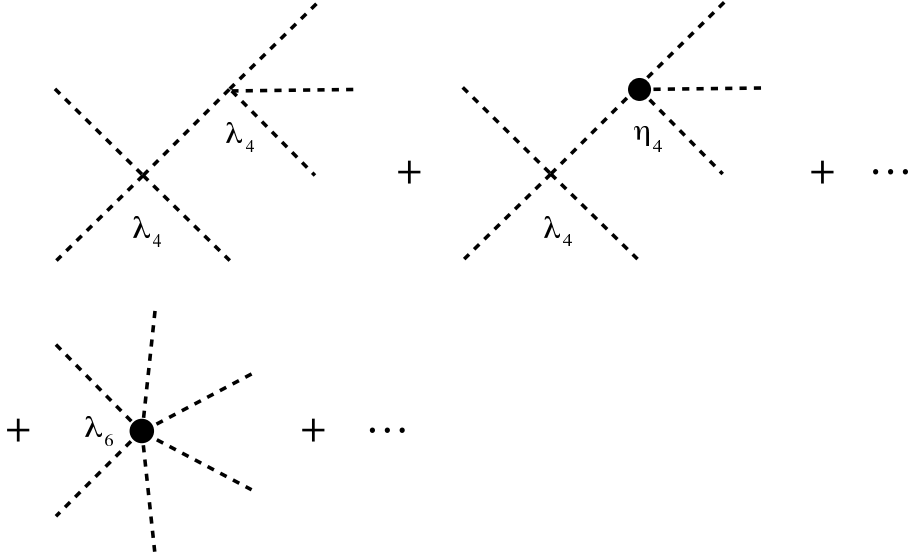}$$
\caption{\label{sixlegs}\em The leading contributions to the inelastic process $2\to 4$.}
\end{figure}

In order to make these general statements more concrete, let us focus on a very simple example. Let us consider
a physical system which at low enough energy possesses just one scalar degree of freedom
 parametrized by a field $\phi$. The most general local and Poincar\'e invariant Lagrangian can be written as an 
 expansion in powers of $\phi$ and of its derivatives
\bea
{\cal L}&=&\partial_\mu\phi\partial^\mu\phi-m^2\phi^2 +\lambda_4\phi^4 +\frac{\lambda_6}{M^2}\phi^6+\frac{\lambda_8}{M^4}\phi^8+\cdots\cr
&+&\frac{\eta_4}{M^2}\phi^2 \partial_\mu\phi\partial^\mu\phi+
\frac{\eta_6}{M^4}\phi^4\partial_\mu\phi\partial^\mu\phi\cdots
\label{lagrangianphi}
\eea
where for simplicity we have also assumed a symmetry $\phi\to -\phi$. We have scaled all the couplings by powers of one mass
scale $M$ and by dimensionless quantities $\lambda_i,\eta_i,\dots$. It is reasonable to assume that  $\lambda_i,\eta_i,\dots\sim O(1)$.
This corresponds to a  theory that in addition to the particle mass $m$ contains only another physical scale, $M$, associated 
to the interactions.
It is easy to understand the meaning of this expansion when calculating scattering amplitudes at energies $m\ll E\ll M$.
Let us focus on tree level computations first. We shall worry about quantum corrections later.
Neglecting numerical factors and indicating one power of momentum generically by $E$, we have
\beq
{\cal A}_{2\to 2}(E)\sim \lambda_4+\eta_4\frac{E^2}{M^2}\dots
\eeq
for the elastic process $2\to 2$ corresponding to Fig.~(\ref{fourlegs}) and
\beq
{\cal A}_{2\to 4}(E)\sim \frac{1}{E^2}\left \{\lambda_4^2+\lambda_4\eta_4\frac{E^2}{M^2}+\lambda_6\frac{E^2}{M^2}+\dots\right \}
\eeq
for the inelastic process $2\to 4$ shown in Fig.~(\ref{sixlegs}). This power counting corresponds to simple dimensional analysis.
Notice that for $E\ll M$ the dimensionless coupling $\lambda_4$ dominates all the amplitudes. This result is intuitively obvious.
A coupling $g$ of mass dimension $[E]^{d}$ can perturbatively contribute to observables via the dimensionless combination
$g/E^d$. We can then distinguish tree classes of couplings depending on whether $d$ is positive, zero or negative.
Couplings of positive dimension are called relevant, as their effect becomes more important the smaller the energy.
An example is given by the mass term itself, which gives small  $O(m^2/E^2)$ effects in the relativistic
regime, but becomes  important when $E\sim O(m)$.
Couplings of vanishing dimensions, like $\lambda_4$, are termed marginal. At tree level their effects are independent of the energy
scale. Finally, couplings of negative dimension are termed irrelevant, as their effects become very small in the low energy domain.
 Notice that while there is only a finite numer of relevant and marginal couplings, the tower of irrelevant couplings is infinite.
In spite of their infinity, and as their naming suggests, irrelevant couplings do not
totally eliminate the predictive power of our
Lagrangian as long as we use it at low energy, $E\ll M$. At each finite order $(E/M)^n$, 
only a finite number
of terms in the Lagrangian contributes to the amplitudes. This preserves a weak form of predictivity, which is often good enough,
since we just need to match our theoretical computations to the experimental precision, which is always finite. 

%
We can now worry about quantum corrections. These introduce some technical difficulties, but the basic conclusion is
unchanged. To be fully general, let us write our Lagrangian as a sum over operators ${\cal O}_i$ of dimension $d_i+4$
\beq
{\cal L}=\sum_i c_i \frac{{\cal O}_i}{M^{d_i}}.
\label{generalop}
\eeq
Assume we want to calculate some observable at order $(E/M)^n$. Working at tree level it is enough to truncate
${\cal L}$ to the operators with $d_i\leq n$. The analysis at tree level is made simple by the fact that the
external momenta ($\sim E$) completely fix the momenta of the internal lines and vertices. This is no longer
true at loop level, where the loop momentum can be arbitrarily high. Moreover some of the loop integrals are UV divergent
and must be cut-off at some scale $\Lambda$.  An interaction  term $c {\cal O}_N/M^{d_N}$, can
generate quantum corrections that involve positive powers of the cut-off $\Lambda$
\beq
\frac{\delta {\cal A}}{\cal A} \sim \cdots +c\frac{\Lambda^P E^{d_N-P}}{M^{d_N}}+\cdots
\label{smallloop}
\eeq
Then an operator with $d_N>n$, which at tree level only gives corrections beyond the needed accuracy $E^n$,
can, at loop level, generate effects that scale with a power $d_N-P\leq n$. Moreover if $\Lambda \sim O(M)$
these quantum effects are as important as the tree level contribution of operators of lower dimension.
This seems very embarrassing. Fortunately it can be proven that these effects are exactly
equivalent to a renormalizations of the coefficients $c_i$ of the operators of lower dimensionality. Therefore they do not contain any new
information and can be eliminated by a trivial change of renormalization scheme.
Their equivalence to local operators is qualitatively understandable: loops of high virtuality 
are small in position space, corresponding to a region of size $1/\Lambda$, and look pointlike with
respect to the long wavelength $1/E$ of the external particles. Another way to understand this result is to take a Wilsonian view
point where $\Lambda$ is the running cut-off. After running down to a scale $\Lambda$ such that $E\lsim \Lambda\ll M$,
the troublesome virtual effects becomes manifestly small: the big effect has been replaced by a local renormalization
of the classical Lagrangian \footnote{To make this argument fully rigorous one should take into account that 
the Wilsonian Lagrangian at scale $\Lambda$ now also contains terms that scale like inverse powers of $\Lambda$. 
Terms proportional to $1/(\Lambda^RM^{S})$ are however fixed  by the renormalized couplings associated to operators
with $d_i\leq S$ \cite{polchinskirg}.}. 
But there is no doubt that the most convenient method to define effective field theories at the quantum level is by
Dimensional Regularization (DR). Dimensionally regulated loop integrals exhibit no powerlike divergences, only logarithmic divergences survive.
The issue we just worried about does not even arise!  The naive power counting we found at tree level  carries over
to the quantum theory up to mild logarithimic corrections. 
\begin{figure}
\begin{center}
\includegraphics[width=6cm, height=4cm]{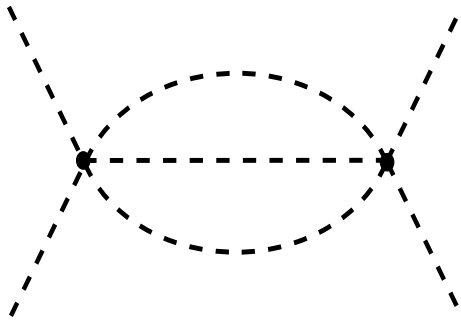}
\end{center}
\caption{2-loop contribution to elastic scattering from $\lambda_5\phi^5/M$ vertices.\label{loop}}
\end{figure}

Consider, for instance,  the 2-loop diagram involving two insertions
of the $\lambda_5\phi^5/M$ interaction shown in Fig.~(\ref{loop}) By using a hard momentum cut off we have
\beq
\delta{\A}_{2\to 2}= \lambda_5^2\left \{a \frac{\Lambda^2}{M^2}+b\frac{E^2}{M^2}\ln \Lambda/E\right\}
\eeq
while  DR in $4+\epsilon$ dimensions gives
\beq
\delta{\A}_{2\to 2}= \lambda_5^2\left (\frac{1}{\epsilon}+b\ln \mu /E\right )\frac{E^2}{M^2}.
\eeq
In DR, after renormalization, this diagram gives just a logarithimic Renormalization Group (RG) evolution of 
the coefficient of a dimension 6 operator
 $(\phi^2)\Dal (\phi^2)$. We emphasize that while the power divergences are totally scheme dependent, being fully saturated in the UV,
the logarithmic divergence involves a physical IR singularity $\ln E$ and must be the same in both regularizations. This $\ln E$ term 
is associated by unitarity to the cut diagrams $|\A_{2\to 3}|^2$.

A by-product of this discussion is that in DR with minimal subtraction 
(or any other mass independent subraction scheme)  the RG equation for the couplings of  an effective Lagrangian 
follows just by dimensional
analysis \cite{Weinberg:1980wa}.  Using the notation of eq.~(\ref{generalop}) 
where  a coupling $c_i/M^{d_i}$ has dimension $-d_i$, the $\beta$ function has the form
\beq
\mu\frac{d c_i}{d \mu}=\sum_{d_j+d_k=d_i} a_{j,k} c_{j}c_{k} +\sum_{d_j+d_k+d_l=d_i} a_{j,k,l} c_{j}c_{k}c_{l} +\cdots
\eeq
where $a_{i,j},\,a_{i,k,l},\, \dots$ are numerical coefficient following from the loop integrals. 
Notice that the parameter submanifold where all the irrelevant couplings  ($-d_i<0$) vanish is stable under RG evolution.
This interesting submanifold corresponds to what we normally call renormalizable theories. On the other hand, once we turn on
an irrelevant coupling of dimension $-d<0$ it will generate by RG evolution an infinite subset of the couplings of
more negative dimension. Such theories are termed non-renormalizable,
as quantum effects force the presence of infinitely many inputs, though we hope to have made it clear how to 
deal with them.
Notice also that our original assumption to scale all the irrelevant couplings by the same mass $M$ is stable under
RG flow. Of course there can be more complicated
situations and models where the higher dimensional couplings involve hierarchically different scales.

We conclude this discussion by reiterating the basic theorem. 
Lagrangians involving all possible non-renormalizable terms
can be made sense of as effective ones. A weak form of predictivity can be preserved by working in perturbation theory
in an expansion in $E/M$, where $M$ is the lowest scale characterizing the non-renormalizable couplings.
  This works as long as $E\ll M$. When $E\sim M$ infinitely many parameters become relevant and our
effective Lagrangian completely looses predictive power. A reasonable expectation is that at the scale $M$ the theory enters
a new regime where perhaps new degrees of freedom are relevant. For instance this is what happens in QCD at the scale $4 \pi f_\pi\sim 1 GeV$
where the weakly coupled description of mesonic physics breaks down. At this energy the hadrons deconfine and at
higher energies the dynamics is more accurately described in terms of quarks and gluons.
Notice that, according to our discussion, non-asymptotically free renormalizable theories are qualitatively
similar to non-renormalizable theories. In the former case, at least 
one coupling grows logarithmically $\lambda(E)\propto 1/\ln(M_L/E)$  with energy, while in the latter the growth
of the effective dimensionless couplings is
powerlike $\lambda(E)\propto (E/M)^n$. In both cases the perturbative description breaks down at some high 
scale, $M_L$ or $M$. 
The only difference between the two cases is  quantitative: in the renormalizable
case, for a not too small a value of  the running coupling at low energy, the cut-off scale $M_L$ is exponentially
far away.   

\subsection{ Examples}
\label{examples}

We can analyze from the effective field theory viewpoint some system of interest. One instructive example is
provided by pure gravity, whose Lagrangian was given in eq. (\ref{Lgravity}). To study the interactions let us
focus on the case of vanishing cosmological constant and let us expand the metric field around the flat background
\beq
g_{MN}=\eta_{MN}+\frac{h_{MN}}{M_D^{\frac{D}{2}-1}}.
\eeq
Eq. (\ref{Lgravity}) will then be written as a power series in the fluctuation $h$
\bea
\left \{ (\partial h)^2 +\frac{1}{M_D^{\frac{D}{2}-1}} h (\partial h)^2+\frac{1}{M_D^{D-2}} h^2 
(\partial h)^2+\dots\right \}\cdots&\cr
+b\left \{
\frac{1}{M_D^{\frac{D}{2}+1}}  (\partial^2 h)(\partial^2 h)+\dots\right \}+ \dots
\eea
where we have been very schematic, suppressing all the tensor structure indices and $O(1)$ factors, but keeping the
derivative expansion structure manifest. Notice that the fluctuation $h$ has been defined in such a way that it is canonically
normalized. The interactions then all scale by inverse powers of the Planck mass. The above Lagrangian, after suitable gauge fixing
can be used to compute graviton scattering processes in perturbation theory. For instance the amplitude ${\cal A}_{2\to 2}$ scales with 
the energy $E$ like
\beq
{\cal A}_{2 \to 2}\sim \frac{E^{D-2}}{M_D^{D-2}}\left (1 + b \frac{E^2}{M_D^2}+\dots\right ).
\eeq
The dots also include quantum effects, which scale like positive powers of $E/M_D$. For instance, by simple dimensional analysis,
 1-loop effects induced by the leading two derivative Lagrangian are 
of order $(E/M_D)^{D-2}$ with respect to the leading tree level contribution. In the ordinary purely 4
dimensional theory of Einstein gravity the Planck mass 
$M_4\sim 10^{18}$ GeV is much bigger than any energy scale relevant to astrophysics or cosmology (if not for very early cosmology, 
even before inflation). Then the truncation
of the theory to the lowest two derivative Lagrangian, the Einstein-Hilbert action, already allows a very
accurate description of the dynamics. 

As a second example consider the interactions of the branon excitations $Y$ of a 3-brane
\beq
{{\cal L}}_{brane}= -f^4 \sqrt{\hat g}+ bf^2 \sqrt{\hat g} \R(\hat g)+\dots
\eeq
Substituting eq. (\ref{Ymetric}) and writing the interactions in terms of the canonical fields
${\hat Y}_i =f^2 Y_i$ it is straighforward to power count the scaling of Feynman diagrams.
For the $YY \to YY$ amplitude the Feynman diagram expansion corresponds to the series
\beq
{\cal A}_{2 \to 2}= \frac{E^4}{f^4}\left (1+ C\frac{E^4}{f^4}\ln E + b\frac{E^2}{f^2}+\dots \right ).
\eeq
where the first and second term (proportional to a calculable coefficient $C$) are determined by the quartic interaction in
eq.~(\ref{tensionexp}) respectively at tree level and at 1-loop.
The tension $f$ turns out to be the energy scale which controls the perturbative expansion.
At energies $E\sim f$ the effective field theory description surely breaks down,  in analogy with the 
case $E\sim M$ in the scalar toy model of the previous section.  The quantity $(E/f)^2$ controls the strength of the 
interaction like $\alpha/4 \pi$ does in quantum electrodynamics. 

The length $L= 1/f$ can be interpreted as the quantum size of the brane, in analogy with the Compton 
wavelength of a particle. Indeed in  the case of a 0-brane, 
a point particle,   $f$ coincides with the mass $m$ and we recover the usual definition of Compton wavelength. 
In the case of a particle the length $1/m$ controls the domain of validity of the low energy non-relativistic
effective theory. If we try to localize one electron at a distance $<1/m$, then, by the indetermination principle,
 not only  will its momentum $p$ be
relativistic but the production of electron positron pairs energetically possible. 
In the case of the brane, we can, for instance, consider the quantum fluctuation  
of the linearized induced metric on the vacuum .
We find (cfr. eq.~(\ref{Ymetric}))
\beq
\langle \partial_\mu Y^i\partial^\mu Y_i\rangle \sim \frac{1}{f^4}\int k^3 dk
\eeq
which shows that at wavelengths of order $1/f$ the fluctuation of the brane position becomes itself of order $1/f$: at these
short wavelengths the brane cannot be approximated by a smooth surface. Basically it is not possible to talk
about fluctuations in the position of the brane that are shorter than $1/f$ in both longitudinal
and transverse directions.

With the previous considerations in mind, it is instructive to consider the field theoretic domain wall discussed 
in sect. \ref{Gravity and Branes}. For definiteness let us focus on the case of a 
4-dimensional scalar theory, so that the wall is a 2-brane. The tension
is $\tau={\sqrt g} v^3$, while the cut-off of the effective  description is provided by $m={\sqrt g} v$, the energy
at which extra massive modes come in. 
We have $m=(g\tau)^{1/3}$, so that as long as the original 4D theory was weakly
coupled ($g\ll 1$), the brane theory never gets into a strong coupling regime. At the cut off scale $m$, the loop 
expansion parameter of the effective brane theory $m^3/\tau=g$ coincides with the loop expansion parameter of 
the original scalar field theory.
 
\begin{figure}
\begin{center}
\includegraphics[width=5cm, height=6cm]{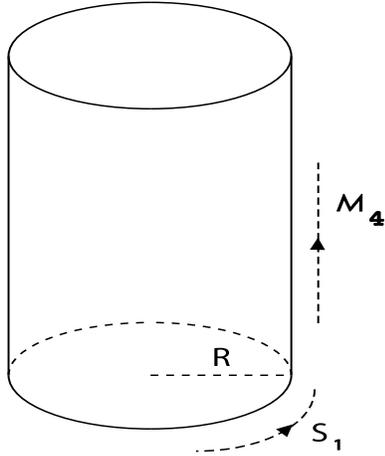}
\end{center}
\caption{Cylindrical structure of 5-dimensional space-time compactified on $\M_4\times S_1$.\label{kk}}
\end{figure}

\subsection{Kaluza-Klein decomposition}
So far we have been general: our discussion applies equally well to compact and to infinite extra dimensions. However, since 
it is empirically
very clear that we live in three macroscopic spatial dimensions, for phenomenological applications we must focus on
the case in which the extra-dimensions are compactified at  some small enough radius $R$. The dynamics at distances
much bigger than $R$ will not be able to notice the presence of the extra compact directions. To illustrate this fact let us consider the
simplest situation of a 5D scalar field $\phi$  with the  5th dimension compactified on a circle (see Fig.~(\ref{kk})) of
 radius $R$. Compactification is formally expressed by  the periodicity requirement
\beq
\phi(x,x_5) = \phi(x,x_5+2\pi R)
\eeq
Processes taking place on time scales $T \ll R$, by causality and by locality, cannot notice that the 5th dimension is compact.
On the other hand to study processes happening on a time scale $T \gsim R$, and in particular at energies $E \lsim 1/R$, the
 5D local description
is not the most adequate. In this case it is convenient to expand the field $\phi$ in its Fourier components with respect 
to the periodic coordinate $x_5$.
\beq
\phi(x,x_5)=\sum_{n=-\infty}^{n=\infty}\phi_n(x)e^{i\frac{n x_5}{R}}.
\eeq
where the reality of $\phi$ implies $\phi_{-n}(x)=\phi_n(x)^*$.
Notice that each different coefficient $\phi_n$ in this expansion corresponds to a different 4D field. The $\phi_n$ are called
Kaluza-Klein (KK) fields. According to this expansion the 5D kinetic action integrated over $x_5$ becomes
\bea
\int {\cal L}_\phi dx^5 =&-\frac{1}{2}&\int \left[ (\partial_\mu\phi)^2-(\partial_5 \phi)^2\right]=\cr
 &\frac{1}{2}&\sum_{-\infty}^{\infty}\left [-|\partial_\mu \phi_n|^2 +\frac{n^2}{R^2}|\phi_n|^2\right ].
\eea
The original 5D massless field has been decomposed in a tower of Kaluza-Klein scalars $\phi_n$ with mass 
\beq
m_n=n/R.
\eeq
If we work at energy $E$, only a limited number $n\sim ER$ of KK can be produced. In particular,
for $E<1/R$  only the zero mode $\phi_0$ is available. At such low energies the model looks 4-dimensional.
The KK particles appear only virtually, and their effect is reproduced by a suitable set of local operators
involving only the massless 4D fields. 
In the specific example we are considering, the full space-time symmetry is just the 4-dimensional Poincar\'e group times
translations along the fifth direction: $P_4\times U(1)$. The KK particle states represent just the irreducible representations
of this group. In particular the index $n$ represents the charge under the $U(1)$  group of 5D translations: 5D translational
 invariance
shows up in 4D as the conservation of the KK indices $n_i$ summed over the incoming and outgoing particles in a collision.

Along similar lines one can study the KK decomposition of a gauge vector field $A_M$. But rather than discussing
it in detail we go directly to the case of the graviton: the technical issues, associated to gauge invariance, are analogous 
for both  vector and tensor field. So let us consider the original theory of Kaluza and Klein \cite{KaluzaKlein}: 5D Einstein gravity
compactified on $\M_4\times S_1$ with the action
\beq
2 M_5^3 \int_{\M_4\times S_1}{\sqrt g} \R(g)
\eeq 
 We can write the full metric tensor in block form
\beq
g_{MN}(x,x^5)=\left (\matrix{g_{\mu\nu} & g_{\mu 5}\cr g_{5\mu} &g_{55}\cr }\right )=
\left (\matrix{\eta_{\mu\nu}+h_{\mu\nu} & h_{\mu 5}\cr h_{\mu 5} &1+h_{55}\cr }\right ).
\eeq

To work out the spectrum we must compute the quadratic action in the linearized field $h_{MN}$ and then use 
the gauge freedom provided by the linearized 5D diffeomorphisms, $x^M\to x^M+\epsilon^M(x,x^5)$
\beq
h_{MN}\to h_{MN}+\delta h_{MN}= h_{MN}+\partial_N\epsilon_M+\partial_M\epsilon_N.
\eeq
to eliminate the redundant degrees of freedom. Here and in what follows,
working at linear order, indices are raised and lowered using the Lorentz metric $\eta_{MN}$. We stress that
the compactification of the fifth dimension implies that all our fields, including $\epsilon_M$ are periodic in $x^5$.
Using the 5 gauge parameters $\epsilon_N$ we can essentially eliminate 5 combinations of the metric 
fluctuations $h_{MN}$. We can choose these 5 combinations to be just $h_{55}$ and $h_{\mu 5}$.
By using Fourier modes we have that
$\delta h_{55}=2 \partial_5 \epsilon_5$ becomes 
\beq
\delta h_{55}^{(n)}=2 i n \epsilon_5^{(n)}
\eeq
which explicitly shows that we can eliminate all the modes but  $h^{(0)}_{55}$, which is gauge invariant.
The gauge invariance of zero modes like $h^{(0)}_{55}$ follows from the periodicity of the gauge transformation
\beq
\oint_{S_1}\delta  h_{55} =2\oint_{S_1}  \partial_5\epsilon_5=0
\eeq
and is a generic features of gauge theories on compact spaces. The same thing happens  for $h_{\mu 5}$
\beq
\delta h_{\mu 5}^{(n)}=\partial_\mu \epsilon_5^{(n)}+ i n \epsilon_\mu^{(n)}.
\eeq
Therefore by using the $n\not = 0$ modes of $\epsilon_M$ we can go to a gauge where
\beq
h_{55}(x,x_5)\equiv \phi(x) \quad\quad h_{\mu 5}(x,x_5)\equiv A_\mu(x)
\eeq
 while $h_{\mu\nu}(x,x_5)$ is unconstrained. However we still have the zero mode
gauge freedom
\beq
A_\mu \to A_\mu+\partial_\mu \epsilon^{(0)} \quad\quad h_{\mu\nu}^{(0)}\to h_{\mu\nu}^{(0)}-\partial_\mu\epsilon_\nu^{(0)}
-\partial_\nu\epsilon^{(0)}_\mu.
\label{residual}
\eeq
The residual scalar mode $\phi$, usually called radion, is associated to fluctutations in the proper length $L$ of the radius of
compactification  $\delta L =\oint h_{55}/2=\pi R \phi$. The graviphoton $A_\mu$, as shown in eq. (\ref{residual}), is
the gauge field associated to 4D local translations of the 5th coordinate. The associated charge is just
the momentum along the fith dimension, i.e. the KK index $n$.  

Defining the KK modes via
\beq
h_{\mu\nu} \equiv \sum_{n=-\infty}^{+\infty}h_{\mu\nu}^{(n)} e^{i\frac{n x_5}{R}}
\eeq
the linearized 4D action becomes 
\bea
 {\cal L}^{(2)}_{4D}=M_5^3\pi R\Biggl \{\Bigl [&\sum_{-\infty}^{+\infty} {h^{(n)}}^{\mu\nu}\Dal h_{\mu\nu}^{(-n)}
 -{h^{(n)}}_\mu^\mu \Dal {h^{(-n)}}_\nu^\nu \cr
+&2 h^{(n)}_{\mu\nu}\partial^\mu\partial^\nu {h^{(-n)}}_\rho^\rho
-2 h^{(n)}_{\mu\nu}\partial^\mu\partial^\rho {h^{(-n)}}_\rho^\nu\cr
+&\frac{n^2}{4R^2}\bigl({h^{(n)}}_\mu^\mu
{h^{(-n)}}_\nu^\nu
 -{h^{(n)}}^{\mu\nu}h_{\mu\nu}^{(-n)}\bigr )\Bigr ]\cr
+&2\phi\bigl (\partial^\mu\partial^\nu h_{\mu\nu}^{(0)}-\Dal {h^{(0)}}^\mu_\mu\bigr )- F_{\mu\nu}F^{\mu\nu}\Biggr \}
 \label{linearizedKK}
 \eea
 where $F_{\mu\nu}=\partial_\mu A_\nu-\partial_\nu A_\mu$. By looking at the coefficient of the zero mode action we deduce that
the effective 4-dimensional Planck scale is
\beq
M_4^2 \equiv M_5^3 2\pi R
\eeq
The $\partial_5$ terms of the 5d Lagrangian have turned into mass terms for the
 $n\not = 0$ modes. As first noticed by Fierz and Pauli \cite{Fierz:1939ix}, the specific tensor structure of this mass term is the only one ensuring 
the absence of ghosts and tachyons in $h^{(n)}_{\mu\nu}$. The equations of motion for the massive modes reduce indeed to
\beq
 \left (\Dal  + \frac{n^2}{R^2}\right ) h^{(n)}_{\mu\nu}=0\quad\quad \quad\partial^\mu h^{(n)}_{\mu\nu}={h^{(n)}}_\mu^\mu=0
\label{onshellKK}
 \eeq
 where the second and third relations follow by taking the divergence and trace of the equation of motion. This is completely analogous to the well 
known case of a massive vector $V_\mu$. There the divergence of the equation of motion gives the constraint 
 $\partial^\mu V_\mu=0$, 
  implying that only  3 out of the 4 degrees of freedom propagate, as it
  should be for a $J=1$ massive particle. Here, due to the constraints, we have $10-5=5$
  propagating states, corresponding to a massive $J=2$ particle.  An arbitrary symmetric two index tensor $H_{\mu\nu}$ can be decomposed
in components of definite spin as
\beq
H_{\mu\nu}= H^{TT}_{\mu\nu}+\partial_\mu E_\nu^T+\partial_\nu E_\mu^T+ \left (\eta_{\mu\nu}-\frac{\partial_\mu\partial_\nu}{\partial^2}
\right )\Phi +\frac{\partial_\mu\partial_\nu}{\partial^2} \Psi\equiv 2\oplus 1\oplus 0_\Phi\oplus0_\Psi
\label{spindec}
\eeq
where $E^T_\mu$ and $H^{TT}_{\mu\nu}$ are respectively transverse and transverse-traceless
($\partial^\mu E_\mu^T= \partial^\mu H_{\mu\nu}^{TT}=\eta^{\mu\nu} H^{TT}_{\mu\nu}=0$) and the spin of each component is indicated 
in an obvious notation. By eq. (\ref{onshellKK}) only $H^{TT}$ survives on-shell. An instructive exercise is to construct the projectors on
 $H^{TT}, E^{T}, \Phi$ and $\Psi$ by writing them in a compact way in terms of the transverse and longitudinal vector projectors 
$\Pi_{\mu\nu}^T=\eta_{\mu\nu}-\partial_\mu
\partial_\nu/\partial^2$, $\Pi_{\mu\nu}^L=\partial_\mu\partial_\nu/\partial^2$.
Another instructive exercise is to write  the kinetic Lagrangian in terms of the various projectors,
in complete analogy with the massive $J=1$ case, and from that to derive the massive $J=2$ propagator
\beq
\langle h^{(-n_)}_{\mu\nu} h^{(n)}_{\rho\sigma}\rangle =\frac{\frac{1}{2}\left( \hat\Pi_{\mu\rho}^T\hat\Pi_{\nu\sigma}^T+
\hat\Pi_{\mu\sigma}^T\hat\Pi_{\nu\rho}^T\right )-\frac{1}{3} \hat\Pi_{\mu\nu}^T\hat\Pi_{\rho\sigma}^T}{p^2+\frac{n^2}{R^2}}
\equiv \frac{\Pi_{\mu\nu\rho\sigma}(m_n)}{p^2+m_n^2}
\label{massiveprop}
\eeq
where $\hat\Pi_{\mu\nu}^T=\eta_{\mu\nu}-p_\mu p_\nu/(m_n)^2$.

Let us now focus on the zero modes. Notice that the radion mixes kinetically to the graviton. It is convenient
to diagonalize the kinetic term via the Weyl shift $h_{\mu\nu}^{(0)}\equiv {\bar h}_{\mu\nu}- \frac{1}{2}\phi\eta_{\mu\nu}$, after which
$\phi$ acquires a self kinetic term
\beq
\frac{3}{2} M_5^3\pi R \phi\Dal \phi
\eeq
while ${\bar h}_{\mu\nu}$ has obviously the kinetic term of massless graviton. At this point we can gauge fix the residual 
4D reparametrization and gauge symmetry by using respectively the de Donder and Feynman gauges
\beq
2M_5^3\pi R\Bigl \{ \left (\partial^\mu {\bar h}_{\mu\nu}-\frac{1}{2} \partial_\nu {\bar h}^\mu_\mu\right )^2-\left (\partial^\mu A_\mu
\right )^2\Bigr \}.
\eeq
On shell we have 2 physical helicity  states in both ${\bar h}$ and $A$. These, including $\phi$, add up to 5 
states: the same number  we found at each excited level, but here they are shared among particles of different spin.

The presence of the radion $\phi$ makes this theory quite different from ordinary 4D Einstein gravity
(the additional scalar is sometimes called a Brans-Dicke field). The tensor field that
couples to ordinary 4D matter and thus describes the observable fluctuations of the 4D geometry is the original metric
$h_{\mu\nu}^{(0)}$ and not ${\bar h}_{\mu\nu}$.  Therefore the relevant graviton propagator is
\bea
\langle h_{\mu\nu}^{(0)} h_{\rho\sigma}^{(0)}\rangle& =&\langle {\bar h}_{\mu\nu} {\bar h}_{\rho\sigma}\rangle
+\frac{1}{4}\langle \phi \phi\rangle\cr
&=&\frac{1}{M_4^2}\bigl \{\frac{\frac{1}{2}\left (\etamr\etasn+\etams\etarn\right )-\frac{1}{2}\etamn\etars}{q^2}\cr
&+&\frac{1}{6}
\frac{\etamn\etars}{q^2}\bigr \}
\label{masslessprop}
\eea
where the first term is just the result we would get in ordinary GR and the second contribution, 
proportional to $1/6$, 
is due to the radion. In the non-relativistic regime the effects of the tensor and scalar field are indistinguishable.
The Newton constant is determined by  $\langle h_{00}^{(0)} h_{00}^{(0)}\rangle$ and given by
\beq
32 \pi G_N=\frac{2}{M_4^2}\left( \frac{1}{2} +\frac{1}{6}\right )=\frac{4}{3}\frac{1}{M_4^2}
\label{Gnewton}
\eeq
where we have indicated separately the contributions of the tensor and of the scalar.
However in the relativistic regime the implications of the two terms are quite different. In particular $\phi$
does not couple to  photons as they have a traceless energy momentum tensor. Now, one of the most accurate tests
of GR is the measurement of the deflection of light by the gravitational field of the Sun: the experimental result
agrees with the theory to about 1 part in $10^{3}$. In the theory at hand, $\phi$ does not contribute to this deflection,
and the scattering angle, expressend in terms of the non-relativistic $G_N$ of eq. (\ref{Gnewton}), is therefore 
only $3/4$
of the GR prediction. This result is completely ruled out by the data. In order to meet consistency, 
$\phi$ should be given a
 mass $m_\phi$, so that its contribution to the potential will decay as $e^{-m_\phi r}/r$ and become 
quickly irrelevant for 
$r>1/m_\phi$. Notice that giving a mass to $\phi$ corresponds to stabilizing the size of the 5th dimension.
 The agreement between the value of $G_N$ measured in
post Cavendish experiments \cite{Adelberger:2002ic,kapitulnik,long} down to distances of order 100 $\mu m$ with that governing post Newtonian corrections 
in the solar system
forces the mass of $\phi$ to be bigger that $\hbar c/100 \mu m\sim 10^{-3}$ eV.


To gain another viewpoint on  the compactification of gravity from $D$ down to 4 dimensions,
it is worth to count the physical states of gravity around ordinary  D-dimensional (non-compact) Minkowsky space.
The metric symmetric tensor $h_{MN}$ corresponds to $D(D-1)/2$ fields. By using the $D$ gauge transformations
$\epsilon_N$ we can eliminate $D$ of these fields. For instance we can go to the Gaussian normal gauge
where all time components vanish $h_{00}=h_{0i}=0$. This choice is the analogue of temporal gauge in Maxwell's theory.
However as in any gauge theory, even after fixing the gauge, we must still impose the equations of motion
of $h_{00}$ and $h_{0i}$ as an  initial time constraint
\beq
\frac{\partial {\cal L}}{\delta h_{00}}= \nabla^i\nabla^j h_{ij}-\nabla^i\nabla_i h^j_j|_{t=0}=0\quad\quad
\frac{\partial {\cal L}}{\delta h_{0i}}= \left (\nabla^j {\dot h}_{ij}-\nabla_i {\dot h}^j_j\right )|_{t=0}=0
\label{initialconstraints}
\eeq
where by the dot and by $\nabla_i$ we indicate respectively detivatives with respect to the time, $t$, and space,
$x^i$, coordinates. 
The divergence of the $h_{ij}$ equations of motion 
\beq
\left (\nabla^j {\ddot h}_{ij}-\nabla_i {\ddot h}^j_j\right )=0
\label{twodots}
\eeq
ensures that both constraints remain valid at all times\footnote{Again this is analogous to electromagnetism. 
In temporal gauge,
the $A_0$ equation of motion constraint gives Gauss law ${\vec \nabla}\cdot {\vec E}=0$ at initial times. Maxwell's 
equation
${\dot {\vec E}}= \vec \nabla \times\vec B $ implies the validity of Gauss's law at all times.}. To clarify things
it  is convenient to decompose $h_{ij}$ in spin components as previously done in eq.~(\ref{spindec})
\beq
h_{ij}\equiv H_{ij}^{TT}+\nabla_i V^T_j+\nabla_j V^T_i +\left (\delta_{ij}-\frac{\nabla_i\nabla_j}{\nabla^2}\right )
H
+\frac{\nabla_i\nabla_j}{\nabla^2} V
\eeq
Moreover it should be noticed that there is a residual gauge freedom preserving the Gaussian normal condition
\beq
\epsilon_0\equiv \epsilon_0(x_i) \quad\quad \epsilon_i\equiv \epsilon^T_i(x_i)-\nabla_i(\epsilon_L(x_i) + t 
\epsilon_0(x_i)).
\eeq
Notice that the $\epsilon$'s depend only on the space coordinates and that $\nabla^i\epsilon_i^T=0$.
Now, the divergence and trace of the $h_{ij}$ equation of motion imply
\beq
\nabla_i {\ddot H}= \nabla^2 {\ddot V}_i^T=0 \quad\quad {\ddot V}+(D-2){\ddot H}-(D-3)\nabla^2 H=0.
\label{simple}
\eeq 
Assuming that our fields $F$ vanish fast enough at spacial infinity, $\nabla^2F=0$ implies $F=0$. The first two 
equations then imply
$H=H_0(x_i)+tH_1(x_i)$ and $V_j^T={V_0}_j^T(x_i)+t {V_1}_j^T(x_i)$. The initial time constraints imply however
$H_0=H_1={V_1}_j^T=0$. In turn  eq. (\ref{simple}) implies  $V=V_0(x_i)+tV_1(x)$. At this point we are left with
$D$ functions ${V_0}_i^T, V_0, V_1$
which can be completely eliminated by the residual gauge freedom $\epsilon_i^T, \epsilon_L, \epsilon_0$.
Notice that the initial time constraints eliminate 1 dynamical variable, $H$, plus $D-1$  ``velocities''
 $\dot H, {\dot V}_i^T$. Instead the gauge freedom allows to eliminate $D-1$ variables and 1 velocity. This 
generalizes the situation in electromagnetism where the Gauss constraint eliminates 1 velocity, 
$\nabla^i {\dot A}_i$, while the residual gauge freedom eliminates $\nabla^i A_i$.





The result of all this is that, after going to the $h_{00}=h_{0i}=0$ gauge,
an additional $D$ deegrees of freedom are eliminated ($V^T_i, H, V$) and we are
left with $D(D-1)/2-2D=D(D-3)/2$ propagating fields, corresponding to $H_{ij}^{TT}$.

Before concluding this section we discuss the more general case of Einstein gravity in $D=4+n$ dimensions, with the
$n$ extra dimensions compactified on a square torus $T^n$ \cite{everybody}. Indicating by $i$ and $x^i$ the extra indices 
and coordinates,
$T^n$ is defined by the equivalence relation $x^i\sim x^i +2 \pi R n^i$ with $n^i$ a vector with integer entries. KK levels
are labelled by a vector of integers $(\vec n)_i= n_i$ associated  to the momentum $n_i/R$ along the $T^n$ directions.
The counting of physical degrees of freedom for each massive KK level is shown in Table (\ref{dof}). It is easy to check 
that gauge invariance allows to eliminate the 4D vector $n^i h_{i\mu}$ (4 fields) and the scalars $n^i h_{ij}$ (n fields).
On shell $n-1$ longitudinal components from the remaining $n-1$ vectors and 5 more components from 
$h_{\mu\nu}$  are eliminated. The propagating degrees of freedom are correspondingly $n-1$ massive vectors and 1 massive graviton.
 Thus there finally result $(n+4)(n+1)/2=D(D-3)/2$ physical states, in agreement with our previous
derivation. At the zero mode level there is the same number of degrees of freedom, but they are shared among 1 graviton 
$h_{\mu\nu}^{(0)}$
, $2n$ graviphotons $A^i_\mu$ and a symmetric matrix of 
$n(n+1)/2$ scalars $\phi_{ij}$. The scalars $\phi_{ij}$ are the moduli describing the fluctuations in the shape and size of the torus. In particular
the trace $\phi^i_i$ describes the fluctuations of the torus volume. This field mixes to the 4D graviton leading to a 
propagator with an extra scalar term
\beq
\frac{1}{M_4^2}\bigl \{\frac{\frac{1}{2}\left (\etamr\etasn+\etams\etarn\right )-\frac{1}{2}\etamn\etars}{q^2}+\frac{n}{2n+4}
\frac{\etamn\etars}{q^2}\bigr \}.
\label{propzeromodes}
\eeq
In order to agree with observations, the volume modulus should be stabilized.

\begin{table}
\begin{center}
\begin{tabular}[h]{|c|c|c|c|c|}
\hline
field & original& $(-)$ gauge& $(-)$ eqs. motion & propagating\\ 
      &d.o.f    &fixing      &                   &d.o.f.\\
\hline
($J=2$)  $h_{\mu\nu}$ & $10$ &$ 0$ & $ -5$ & $5$ \\ 
\hline
($J=1$)  $h_{\mu i}$ & $4n$ & $-4$ & $-(n-1)$& $3(n-1)$ \\ 
\hline
($J=0$) $h_{ij}$ & $\frac{n(n+1)}{2}$ & $-n$ & $0$ &$\frac{n(n-1)}{2}$\\ 
\hline
\end{tabular}
\caption{\label{dof} Number of degrees of freedom (d.o.f.) off-shell and on-shell for each field component}
\end{center}
\end{table}

\subsection{Large Extra-Dimensions}
\label{LED}
A very interesting arena where to apply the concepts that we introduced is given by the scenario
of large extra dimensions. This scenario has been advocated by Arkani-Hamed, Dimopoulos and Dvali (ADD)
as an alternative viewpoint on the gauge hierarchy problem \cite{add}. With respect to the standard picture for physics
beyond the SM the ADD proposal represents a dramatic shift of paradigm. In the standard scenario, fundamental
interactions are described by an ordinary quantum field theory up to energy scales larger that
the Grand Unification scale $10^{16}$ GeV. Above this scale quantum gravity effects or string theory imply a radical revision
of fundamental physics. According to the ADD proposal, instead, this radical revision is needed right above the weak scale!
The proposal is specified by three main features
\begin{itemize}
\item There exists a number of $n$ new spacial compact dimensions. For instance a simple manifold could be just 
$\M_4\times T^n$.
\item The fundamental Planck scale of the theory is very low $M_D\sim$ TeV.
\item The SM degrees of freedom are localized on a 3D-brane stretching along the 3 non-compact space dimensions.
\end{itemize}
As we will now explain, these three requirement allow for a drastically different viewpoint on the hierarchy problem,
without leading to any stark disagreement with experimental observations. Let us focus on gravity first. As we have
already seen in the simple case of Kaluza-Klein's theory, the macroscopic Planck mass $M_4^2$ of the effective 4D theory
is related to the microscopic $M_D$ via
\beq
M_4^2= M_D^{2+n} V_n
\label{savas}
\eeq
where $V_n$ is the compactification volume. For a torus we have $V_n=(2\pi R)^n$ and the above result
follows from a simple generalization of the analysis we previously did on $S_1$. We can also obtain this relation
by considering directly the effective action for a purely zero mode $g_{\mu\nu}(x^\mu, x^i)\equiv \bar g_{\mu\nu}(x^\mu)$
fluctuation of the metric along $\M_4$
\beq
2M_D^{2+n}\int d^4x^\mu d^nx^i \sqrt{g}{\cal R}_D(g) \quad \leftrightarrow \quad 2 M_D V_n \int d^4x^\mu \sqrt{\bar g}
\R_4(\bar g) 
\eeq
where we have explicitly indicated the dimensionality of the Ricci tensor. The main remark of ADD is based on eq. (\ref{savas}).
Provided the volume of compactification is large enough, even a low gravity scale $M_D$ can reproduce
the physical value $M_4=2\times 10^{18}$ GeV. Before discussing the needed size of $R$, notice that eq. (\ref{savas}) has a very simple
interpretation via Gauss's theorem. Consider the Newtonian potential $\varphi\equiv h_{00}/2$ 
generated by a test mass $M$ in the linearized approximation.
At a distance $r\ll R$ the
 compactness of the extra dimensions does not play a relevant role: the potential is to a good approximation 
 $SO(3+n)$ symmetric and given by
\beq
\varphi |_{r\ll R} \simeq- \frac{\Gamma(\frac{n+3}{2})}{(2n+4)\pi^{\frac{3+n}{2}}}\frac{1}{M_D^{2+n}}\frac{M} {r^{1+n}}.
\label{close}
\eeq
as dictated by Gauss's theorem.
At $r\gg R$ the field lines stretch along the 3 non-compact directions, the 
potential is only $SO(3)$ symmetric. The surface encompassing the field flux is now the two sphere (non-compact directions) 
times the compactification manifold; for instance $S_2\times T^n$. Applying Gauss's theorem we find then
\beq
\varphi |_{r\gg R}\simeq -\frac{n+1}{16 \pi(n+2) M_D^{2+n} V_n}\frac{M}{r}\equiv \frac{n+1}{16\pi(n+2)}\frac{1}{M_4^2}\frac{M}{r}
\label{far}
\eeq
from which we recover again eq. (\ref{savas}). In practice the large distance field is made weaker by the large
extradimensional volume in which the field lines can spread. (The dependence of eq.~(\ref{far}) on $n$ 
is due to the massless radion. For $n=1$ it agrees with eq.~(\ref{Gnewton}), while for general $n$ 
eq.~(\ref{far}) is simply the Fourier transform of eq.~(\ref{propzeromodes}).).

If the ultimate cut-off  $M_D$ is of order the weak scale itself $G_F^{-1/2}$, 
then the expected quantum corrections to the Higgs mass
are of the order of its phenomenologically favored value $m_H\sim G_F^{-1/2}$. In this respect the hierarchy problem, in its ordinary formulation,
is practically eliminated when $M_D\sim 1$ TeV.  With this input, and with the observed value of $M_4$,
 eq. (\ref{savas})  predicts the size of $V_n$. In Table \ref{radii}, we give the radius of compactification in the case of
 a square n-torus. We stress, see eqs. (\ref{close},\ref{far}), that Newton's law is reproduced only at distances larges than $R$. 
\begin{table}
\begin{center}
\begin{tabular}{|c|c|}\hline
$n$ & $R $ \\ \hline\hline
1& $6\times 10^{13}$ cm\\ \hline
2& $ 0.4\,{\rm mm}=1/(10^{-4}{\rm eV})$\\ \hline
4& $ 10^{-8}\,{\rm mm}=1/(20{\rm KeV})$\\ \hline
6& $ 2.5\times 10^{-11}\,{\rm mm}=1/(10{\rm MeV})$\\ \hline
\end{tabular}
\caption{\label{radii} Radius of compactification for fixed value of $\MGRW = 1$ TeV , where $(M_D^{\scriptscriptstyle GRW})^{2+n}
\equiv 4 (2\pi)^n M_D^{2+n}$ is the Planck mass defined in the first paper of ref.~\cite{everybody}.} 
\end{center}
\end{table}
The case $n=1$ requires a radius of compactification of the size of the solar system, which is largely 
ruled out. However already for $n$ greater or equal than 2 the resulting radius is not unreasonable. Indeed experimental tests of gravity at distances shorter than a millimeter are extremely arduous. This is largely due to the presence of  Van der Waals forces,
which tend to swamp any interesting measurement. At present the best bound relegates $O(1)$ deviations from Newton's law
(the ones we would expect is our scenario at $r\lsim R$) to distances shorter than $200 \mu m$. 
In this respect the case $n=2$ is not barely inconsistent. $n=2$ is also experimentally interesting, 
as it predicts deviations in the range  of present sensitivities. The search for deviations from Newton's law is an active
experimental field, also greatly stimulated by the ADD proposal. 

Focusing on gravity only, we have shown  that for $n\geq 2$ the radius of compactification is small enough.
On the other hand the Standard Model as been verified
 down to distances much shorter than the radii shown in the table. 
  The SM is a 3+1 dimensional quantum field theory and its predictions depend crucially on this property.
  LEP, SLC and Tevatron have tested the SM up to an energy of order 1 TeV, corresponding to a  distance
   of order $10^{-16}$ cm. Experimentally then, the SM is a 3+1 dimensional system down to a distance much shorter
    that the radius of compactification. Localizing all the SM degrees of freedom on a 3-brane is an
    elegant way to realize this experimental fact, while keeping  larger radii of compactification. Now it will be the brane size,
    or whatever other characteristic brane cut-off scale, perhaps $1/M_D$ itself,
     to characterize the length scale down to which the SM is a valid 
    effective field theory. This scale can conceivably be $\gsim 1$ TeV.  For instance, the ADD scenario could be realized
    in type I string theory \cite{aadd,ibanez} with the SM localized on a D-brane. In this case the string scale $M_S$, governing the mass of Regge resonances, 
acts as  UV cut off of the brane effective theory.
    
  This completes the basic description of the ADD scenario. It must however be said that, as it stands, 
the ADD proposal is a reformulation of the hierarchy 
problem and not yet a solution \cite{ahdmr}.
 Instead of the small
Higgs vacuum expectation value (VEV) of the old formulation, we now need to explain why the compactification volume $V_n$
is so much bigger that its most natural scale $1/M_D^n$: 
\beq
V_nM_D^n\sim 10 ^{33}.
\eeq
 $V_n$, or equivalently the radius $R$, is a dynamical degree of freedom, a scalar field. We have already shown that in the case of $T^n$
 the fluctuation of $V_n\equiv (2\pi R)^n$ corresponds at linear order to the trace $h^i_i$.
  Since we want a large $\langle R\rangle$  the 
scalar potential $V(R)$ will have to be much flatter than naively expected
 at large values of $R$. As far as we know, the most natural way to achieve
such flat potentials is by invoking supersymmetry. So, if the ADD scenario
is realized in Nature it is likely to be so together with supersymmetry at some stage.
Notice that in the conventional formulation of the hierarchy problem supersymmetry is
invoked to ensure a flat potential at small values of the Higgs field, {\it i.e.} a small Higgs mass. 
As a matter of fact, the ADD proposal maps a small VEV problem into a basically equivalent  large
VEV problem. In the new scenario the hierarchy problem has become a sort of
cosmological constant problem. Indeed a vacuum energy density
$\Lambda^{4+n}$ would add to the radius potential a term $\sim \Lambda^{4+n}R^n$.
This grows very fast at large $R$ so we expect \cite{ahdmr} that $\Lambda^{4+n}$ should be much smaller than
its natural value $(\tev)^{4+n}$. In this respect the presence of bulk supersymmetry would be a natural way to enforce
a small $\Lambda^{4+n}$, thus helping to explain the large volume.
 Indeed ref. \cite{Arkani-Hamed:1999dz} 
presents a simple mechanism which produces large radii at $n=2$, but  which works for a vanishing bulk 
cosmological constant $\Lambda^6$.
Although the model considered  is not supersymmetric it is conceivable that the same mechanism will generalize to 
a supersymmetry set-up and thus lead to a truly natural generation of the hierarchy.

One reason why the ADD proposal is  important is theoretical. The hope is that such a drastic revision of  our view of fundamental
interactions may open the way to new solutions to old problems, like the cosmological constant problem for instance.
Having string theory right at the weak scale may also end up being the right ingredient to build the right string model.
However none of these breakthroughs has come yet. The interest in the ADD proposal is at the moment associated to its
potentially dramatic phenomenological implications \cite{add2}. 
There are two classes of laboratory tests of large extra-dimensions.
We have already commented on the first class, the search for deviations from Newton's law at short but macroscopic distances.
This is done in table top experiments. These deviations could be determined by
the light moduli, like the radius $R$ \cite{ahdmr}, or by the lowest Kaluza-Klein (KK) J=2
modes. Another source of deviation could be the lowest KK mode of a bulk vector field
gauging baryon number \cite{add2}. 
 At present, $O(1)$ deviations from Newton's law have been excluded down to a length $\sim 200\,\mu{\rm m}$ 
\cite{Adelberger:2002ic}, while forces that have a strength $> 10^4$ of gravity are bounded
to have a range smaller than $20\, \mu {\rm m}$ \cite{kapitulnik,long}.
 Notice that this class  of effects crucially depends on the 
features of the compactification manifold at large ``lengths'', as they determine the masses of
 the lightest modes. For instance the presence of even a small curvature of the compactification manifold can
drastically affect these prediction by lifting the lightest states. On dimensional grounds, if the typical curvature length is 
$L$ the modes with mass $<1/L$ will be affected and possibly made heavier.
\begin{figure}
\begin{center}
\includegraphics[width=4cm]{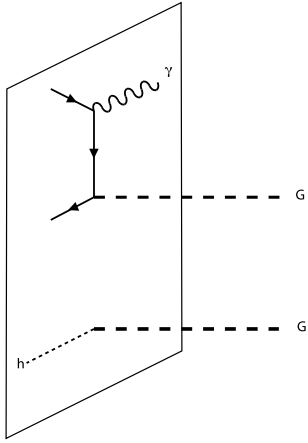}
\end{center}
\caption{Example of two processes with missing energy by bulk graviton radiation.\label{gravemission}}
\end{figure}

The second class of tests is given by high energy collisions \cite{everybody}. 
In this case we deal with either gravitons at virtuality $Q\gg 1/R$ or with real gravitons
measured with too poor an energy resolution to distinguish individual KK
levels. 
 In practice, for high energy processes
happening on a short time scale $\tau\sim 1/E \ll R$, causality and locality imply that we cannot 
notice that the extra dimensions
are compact. Therefore we can take the limit $R\to \infty$ and work as if our brane were embedded
in infinite ($4+n$)-dimensional Minkowski space. (If the compactification manifold had curvature length
$L\ll R$, then the same reasoning would apply for energies $E\gg 1/L$). 
Moreover at energy $E<M_D$, as discussed in section \ref{examples}, we can reliably compute the amplitudes
in a systematic derivative expansion. The characteristic signals are then associated to the emission
of gravitons ($G$) which escape
 undetected into the  extra dimensions.
 Interesting examples (see the figure) are given by the processes
 $e^+e^-\to \gamma+ G=\gamma+\Emiss$ or $pp \to \jet +\Emiss$ or
possibly by the  invisible decay of the Higgs into just one graviton \cite{grw2}. The latter process does not violate
the conservation of angular momentum, since there are KK gravitons of spin 0. 
The relevant interactions are obtained by expanding the brane action for the SM (as in eq. (\ref{braneaction})) in powers of the 
induced metric. For processes with the emission of one graviton we have just the energy momentum term
\beq
{\cal L}_{int} = -\frac{1}{2}T_{\mu\nu}^{SM} h^{\mu\nu}(x,y^i=0)
\eeq
where we have assumed the brane to be located at $y^i=0$. Emission rates can be computed by expanding
$h_{\mu\nu}(x,y_i=0)$ in KK modes.
For instance, the differential cross section for $e^+e^- \to \gamma G$
\bea
\frac{d^2\sigma}{dx_\gamma d\cos \theta}(e^+e^- \to \gamma G)
&=\frac{\alpha}{128}~\frac{1}{(4\pi)^{\frac{n}{2}}\Gamma(\frac{n}{2})}~ 
\left( \frac{{s}^{\frac{n}{2}}}{M_D^{n +2}}\right )\cr
&\left [F_n(x_\gamma, \cos\theta)+O(\frac{s}{M_D^2})\right ]
\label{sezgg}
\eea
\beq
F_n(x,y)=\frac{2(1-x)^{\frac{n}{2} -1}}{x(1-y^2)}
\left[ (2-x)^2(1-x+x^2)-3y^2x^2(1-x)-y^4x^4 \right] .
\label{funz}
%
\eeq
Here $x_\gamma =2E_\gamma /\sqrt{s}$, $E_\gamma$ is the photon energy,
and $\theta$ is the angle between the photon and beam directions. At leading order in $E/M_D$
this process is predicted just in terms of one new parameter, $M_D$ itself. Higher order corrections will depend
on new operator coefficients.  Similar results can be found for all
the other processes \cite{everybody}. 

Notice that graviton emission violates the conservation of momentum along the $y$ directions. This is not surprising
since the presence of the brane at $y=0$ breaks translation  invariance.  
However one might worry that non-conservation of the brane energy momentum might lead to inconsistencies
in the interaction with gravity; after all Einstein equations demand  energy momentum conservation.
 But this is not the case. Even though the {\it global} momentum $P_i$ along
$y_i$ is not conserved (or better not defined) the full energy momentum tensor $T_{MN}$ for the brane plus matter
is indeed {\it locally} conserved. Conservation of a local current in the absence of a globally conserved charge
is the landmark of spontaneous symmetry breaking, and is precisely what happens here. As we have said before,
 translations in  $y^i$
are a non-linearly realized symmetry, with the branons $Y^i$ acting as Goldstone bosons. Local conservation of momentum
 implies the presence of the branons and hence their production in fundamental processes. Since the brane
is infinitely massive it cannot undergo  a {\it global} recoil, but conservation of its energy momentum tensor 
implies the possibility of {\it local} recoil by branon emission. The emission  of branons is another possible signature of
the braneworld scenario. Branons $Y$ are emitted in pairs. At lowest order one finds \cite{Creminelli:2000gh}
\beq
\frac{d^2\sigma}{dx_\gamma d\cos \theta}(e^+e^- \to \gamma YY)
=\frac{\alpha}{1920 \pi^5}~ 
\left( \frac{s^3}{f^8}\right )
\left [F_6(x_\gamma, \cos\theta)+O(\frac{s}{M_D^2})\right ]
\label{branonprod}
\eeq
which, up to an overal constant, is the same result one obtains for graviton emission at $D=10$. A similar result is obtained
in the case of hadronic collisions.
 Comparing 
eqs. (\ref{sezgg},\ref{branonprod}) to experiments one obtains experimental bounds of the scales $M_D$ and $f$. As the effect
grows with $\sqrt s$ the best bounds are obtained from the higher energy experiments, LEP2 and Tevatron.
In particular LEP2 implies the bound $f>100$ GeV \cite{Creminelli:2000gh}. The combined LEP Tevatron bounds on $M_D$ 
\cite{Giudice:2003tu}are shown in Table \ref{leptevatron}.
\begin{table}
\begin{center}
\begin{tabular}{|c|c|c|c|c|c|}\hline
$n$ & $2 $&$3$&$4$&$5$&$6$ \\ \hline\hline
$\MGRW$ (TeV)& $>1.45$ &$>1.09$ &$>0.87$ &$0.72$&$0.65$\\ \hline
\end{tabular}
\end{center}
\caption{\label{leptevatron} Collider bounds on $\MGRW$}
\end{table}
These direct bounds are not very strong. The reason is that the cross section is suppressed by a rather 
large power of $E$ over  the high scale. The LHC will be a better machine to test these ideas through
the direct production of gravitons or string states \cite{everybody,Cullen:2000ef}.
At present stronger bounds come from other effects, associated to contact 4 fermion interactions, that are expected
in the ADD scenario model without being a distinctive feature. The translation of the 
bounds from these other effects into bounds on $M_D$ is a model dependent procedure, but it is fair to say that
they roughly  imply $M_D \gsim 3$ TeV \cite{Giudice:2003tu}. Finally we should mention that, with enough luck, the LHC
may also study gravitational scattering at energies in excess of the Planck mass $M_D$, the so-called transplanckian regime. 
For $M_D\sim 1-3$ TeV, the most energetic collisions at LHC, at $\sqrt s=14$ TeV, should start manifesting the transplanckian
dynamics, which consists of black-hole \cite{savas} or string ball \cite{roberto} production and also of 
the characteristic gravitational elastic scattering \cite{GRW3}. The common features of these processes is to give cross
section at high energy, and fixed angle, that asymptotically grow like a power of energy. This would be an undisputable signal
that the high energy dynamics of gravity, a force whose associated charge is  energy itself, has been detected.

Computations like those we have outlined are relevant also to study the cosmological and astrophysical implications of
the ADD scenario. The phenomenology of these models is now a very wide field. Unfortunately in order to cover it appropriately
we would be lead outside the main goal of the present lectures, which concerns the basic physics principles and tools.
Excellent pedagogical reviews of recent results with extensive references are given by refs. \cite{Rubakov:2001kp,Csaki:2004ay}.

\section{Part II: the Randall-Sundrum model}

In the second part of these lecture we will focus on a specific model proposed by Randall and Sundrum (RS)
which is interesting both from the theoretical and the pedagogical viewpoint. On one side the RS model
realizes a new way of approaching the hierarchy problem, on the other it is simple enough to allow a number of
rather instructive computations. Moreover it naturally demands the introduction of the concept of holography, the basis
of the $AdS$/CFT correspondance \cite{maldacena,witten}. All these reasons make the RS model  a very rich and instructive lecture subject.

\subsection{The Model}

\begin{figure}
\begin{center}
\includegraphics[width=5cm]{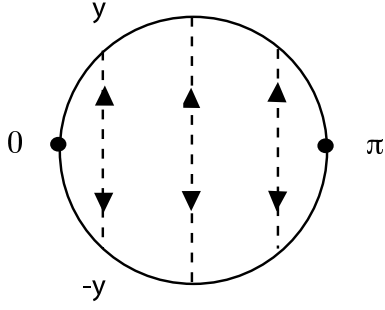}
\end{center}
\caption{Graphical representation of $S_1/Z_2$.\label{orbi}}
\end{figure}

Let us consider a model with a 5th dimension and let us compactify it by considering the following 
equivalence relations for the fifth coordinate $y$
\beq
y \sim y +2 \pi \quad\quad\quad y\sim -y.
\label{equivalence}
\eeq
The first relation, alone, would define a circle $S_1$. The second relation, a $Z_2$ reflection,
 implies identification of opposite points on the circle,
as shown in the Fig.~(\ref{orbi}).  $y=0$ and $y=\pi$ are fixed points under $Z_2$ on the circle and are identified with 
themselves.
The resulting space from this identifications is called a $S_1/Z_2$ orbifold. $S_1/Z_2$ is equivalent to the $[0,\pi]$
segment, but for computational purposes it is useful work with the full $S_1$ covering space with $Z_2$ identification.
Consider now the metric on this space
\beq
ds^2=g_{\mu\nu}dx^\mu dx^\nu +2 g_{\mu 5} dx^\mu dy +g_{55}dy^2.
\eeq
The points related by $y\to -y$ are physically identical, and under $Z_2$ the interval $ds^2$ should be invariant.
Since  $dy \to -dy$ under $Z_2$, we then have
\beq
g_{\mu\nu}(x,y)=g_{\mu\nu}(x,-y)\quad g_{55}(x,y)=g_{55}(x,-y)
\quad g_{\mu 5}(x,y)=-g_{\mu 5}(x,-y).
\label{z2parities}
\eeq
From the last identity and by the continuity of $g_{\mu 5}$ it follows that $g_{\mu 5}\,(x,0)$ $= 
g_{\mu 5}(x,\pi)=0$. Changes of coordinates
\beq
\bar x^\mu ={\bar x}^\mu(x,y) \quad\quad \bar y={\bar y}(x,y)
\label{diff5}
\eeq
should still parametrize an orbifold and respect eq.~(\ref{equivalence}). Without loss of generality we can 
impose $\bar y$ to be the new orbifold coordinate satisfying
\beq
{\bar y}(x,y+2\pi)={\bar y}(x,y)+2\pi \quad\quad {\bar y}(x,-y)=-{\bar y}(x,y)
\eeq
generalizing what we did before for the circle. Notice that under reparame\-trizations the fixed point are mapped
onto themselves ${\bar y}(x,0)=0$, ${\bar y}(x,\pi)=\pi$ \footnote{On the segment $[0,\pi]$
these are the reparametrizations that do not move the boundaries. One could allow more general reparametrizations
under which the boundary points are shifted. The resulting field space would be obviously larger.
The physics would however not be affected. From the point
of view of this more general formulation our field space is obtained just by a partial gauge fixing.}
At the boundaries, eq. (\ref{diff5}) reduces to 4D reparametrizations of the boundaries themselves
\bea
0) &\quad x^\mu \to {\bar x}^\mu(x,0)\cr
\pi) &\quad x^\mu \to {\bar x}^\mu(x,\pi)
\eea
under which the induced metrics
\beq
{g_0}_{\mu\nu}\equiv g_{\mu\nu}(x,0)\quad\quad {g_\pi}_{\mu\nu}\equiv g_{\mu\nu}(x,\pi)
\eeq
are covariant tensors. Now, using $g_{0,\pi}$ we can write the most general invariant action involving also
fields and interactions localized at the boundaries

\bea
S&=&\int d^4x \int_0^{2\pi} dy
\Bigl \{{\sqrt g} \left [2M_5 \R(g)-\tau\right ]\\\nonumber
&+&\delta(y){\sqrt g_0}\left [
{\cal L}_0-\tau_0\right ] +\delta(y-\pi){\sqrt g_\pi}\left [{\cal L}_\pi-\tau_\pi\right ] \Bigr \}
\eea
\noindent
where $\tau$ and  $\tau_0,\tau_\pi$ are respectively the bulk cosmological constant and boundary tensions.
By ${\cal L}_{0,\pi}$ we indicate any other interactions involving fields localized at the boundary.
Neglecting the latter, the 5D Einstein equations are

\bea
&{\sqrt g}\, G_{MN}=\cr 
& \frac{-1}{4 M_5^3}
\Bigl [ \tau {\sqrt g} g_{MN} 
+\left (\tau_0 {\sqrt g_0} {g_0}_{\mu\nu}\delta(y)
+ \tau_\pi {\sqrt g_\pi} {g_0\pi}_{\mu\nu}\delta(y-\pi)\right )
\delta^\mu_M \delta^\nu_N\Bigr ]\cr
\eea

We look for a solution with the following Poincar\'e invariant ansatz 
\beq
ds^2 =e^{-2\sigma(y)}\eta_{\mu\nu}d x^\mu d x^\nu + r_c^2 dy^2
\eeq
where $r_c$ is a parameter with dimension $[{\rm length}]$ parametrizing the proper distance between the two fixed 
points. 
The $\mu 5$ equation is identically satisfied while the $55$
is
\beq
\left (\frac {\sigma^\prime}{r_c}\right )^2=-\frac{\tau}{24 M_5^3}\equiv k^2.
\label{55}
\eeq
implying a negative cosmological constant $\tau<0$.
Imposing the orbifold symmetry  eq.~(\ref{55}) has two solutions (up to trivial coordinate changes):
\beq
\sigma(y)= \pm kr_c|y|.
\label{warpfactor}
\eeq
Finally the $\mu\nu$ equation is
\beq
\sigma''= \frac{r_c}{12 M_5^3}\left [\tau_0\delta(y)+\tau_\pi\delta(y-\pi)\right ]
\label{munu}
\eeq
which is solved by eq. (\ref{warpfactor}) provided the following conditions hold
\beq
\tau_0=-\tau_\pi =\pm 24 M_5^3 k.
\label{tuning}
\eeq
Without loss of generality, we can choose the solution $\sigma=+k r_c |y|$. Notice that the
tension $\tau_\pi$ is negative. We will later show that this does not lead to instabilities.
Eq. (\ref{tuning}) represents a tuning of two parameters, in the absence of which 
there would not exist a static solution with  Poincar\'e symmetry. This
 does not seem at all a desired feature for a model aiming at a solution of  the gauge hierarchy problem!
The meaning of eq. (\ref{tuning}) will become more clear below. We will then explain that in the complete model
there remains just one tuning, the one associated to the 4D cosmological constant. This is a situation
common to all the other solutions of the gauge hierarchy problem, like supersymmetry or technicolor.

Our orbifold with  metric
\beq
ds^2=e^{-2kr_c|y|}dx_\mu d x^\mu+ r_c^2 dy^2
\label{solution}
\eeq
corresponds to two slices of 5-dimensional anti-de Sitter space (AdS5) glued back to back at the fixed points. 
The full AdS space is obtained by $|y|\to y$ and for $y\in (\-\infty,+\infty)$. $k^2=\R_5(g)$ parametrizes the 
curvature of
this space. In order for our  effective field theory to be a valid description it should be $k\ll M_5$. 
A region with fixed $x$ coordinates exported along $y$ describes a throat that gets
exponentially narrow at larger $y$. Fig. 9 depicts the same contraction for the wavelength of infalling
quanta, which we will later discuss in more detail. It is sometimes said that the space is ``warped'' by the $y$ dependent factor
multiplying the 4D metric. 

\begin{figure}
\begin{center}
\includegraphics[width=5cm]{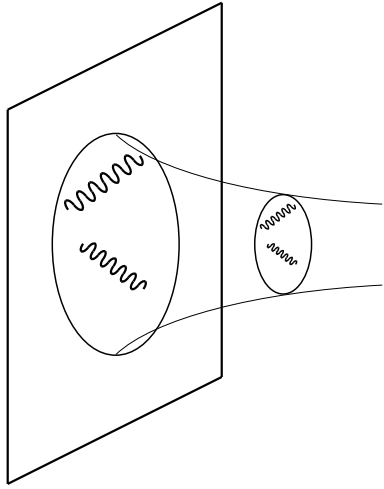}
\caption{\label{blueshift}\em Contraction of wavelengths as particles fall through the $AdS$ throat.}
\end{center}
\end{figure}

 Our metric can be used to Redshift 4-dimensional mass parameters as suggested
at the beginning of these lectures. Assume we place at $0$ and $\pi$ two identical copies of a 4D QFT.
Exactly like with two atoms in the gravitational field of a star, any direct experimental comparison
of the masses of the equivalent states at each brane  gives
\beq
\frac{m_\pi}{m_0}=e^{-k r_c\pi}.
\eeq
For $e^{-kr_c \pi}\sim m_Z/M_4$ this effect could be relevant in explaining the weak scale hierarchy.
Because of the relative shift of mass scales the $0$ and $\pi$ fixed points are called respectively
the Planck and TeV brane.
Notice also that, because of the appearance of an exponential factor, a Redshift of order $m_z/M_4\sim 10^{-16}$ 
is already obtained when
the  proper radius of compactification 
$ r_c \pi$ is only about $35-40$ times larger than the curvature radius $1/k$. The latter could in turn 
be not much bigger than the 5D Planck length $1/M_5$. Therefore a large hierarchy can be obtained from a rather small
fifth dimension. 

\subsection{Low energy effective theory}
The RS model is a generalization of the Kaluza-Klein theory that we already studied. It is then straightforward for 
us 
to discuss its zero mode content and to derive the low energy effective field theory. The main difference
with respect to gravity on $S_1$, is that $g_{\mu 5}$
is odd under orbifold parity,  see eq. (\ref{z2parities}). Then
$g_{\mu 5}$ does not have a graviphoton zero mode and we can go to a gauge where $g_{\mu 5}\equiv 0$. On the other 
hand 
the graviton and radion zero modes are just obtained by promoting the Lorentz metric $\eta_{\mu\nu}$ and radius $r_c$ 
to 4-dimensional fields \cite{rs}
\beq
ds^2=G_{MN}dX^NdX^M= e^{-2k T(x) |y|}\bar g_{\mu\nu}(x) d x^\mu d x^\nu +T(x)^2 dy^2.
\label{fluctuations}
\eeq
Notice indeed that when  $\bg_{\mu\nu}$ and $T$ are constant over space-time, eq. (\ref{fluctuations}) is a solution
of the equations of motion.($T=r_c$ is not fixed by the solution, while the solutions with arbitrary constant $\bg_{\mu\nu}$
is simply obtained from the original solution  by applying  
the coordinate change,  $ x^\mu \to A^\mu_\nu x^\nu$, with $A_\mu^\rho A_{\rho \nu}=\bg_{\mu\nu}$.)
 Then the 5D action is stationary over the field manifold of constant
$\bg_{\mu\nu}$ and $T$, {\it i.e.} there is no ``potential'' for $\bg_{\mu\nu}$ and $T$, so that they must 
correspond to massless particles. 
 By substituting the ansatz in eq.~(\ref{fluctuations}) in the action we find indeed
\beq
{\cal L}_4^{eff}=\int {\cal L}_5(\bg,T) dy =\left (\frac{M_5}{k}\right )^3 \sqrt{ \bg}\Bigl \{ \left (2k^2 -2\mu^2\right )\R(\bg)
+12 (\partial \mu)^2\Bigr \}
\label{effectiveLradion}
\eeq
where
\beq
\mu(x) = k e^{-k T(x)\pi}.
\label{radion}
\eeq
and where all the metric contractions are done with $\bg_{\mu\nu}$.
This Lagrangian correctly describes the interactions of the zero modes up to terms with more than two derivatives.
We will explain this in more detail when deriving the KK spectrum. By substituting $\langle T\rangle =r_c$ in the 
above equations
we can read the effective 4-dimensional Planck scale
\beq
M_4^2=\frac{M_5^3}{k}\left (1-e^{-2kr_c \pi}\right ).
\label{4dplanckmass}
\eeq
By working with $T\equiv r_c$, eq. (\ref{fluctuations}) substituted in the 5D Einstein term gives 
$\sqrt{g}\R_5(g)=\sqrt{\bg} r_c e^{-2\sigma(y)}\R_4(\bg)+\dots$.
Eq.~(\ref{4dplanckmass}) then simply corresponds to the integral
\beq
M_4^2=M_5^3 \int_{-\pi}^\pi e^{-2\sigma(y)} r_c dy.
\eeq
The crucial aspect of this result is that  $M_4$ is completely dominated by the region close to the Planck brane,
where the warp factor is of order 1. Therefore the value of $M_4$ is  insensitive to the Redshift of mass scales
that takes place in the bulk and which is maximal at the TeV brane. 

Now that we have calculated the Planck mass we can discuss  the issue of the gauge hierarchy in more detail. 
Assume we localize all the fields
of the Standard Model on the TeV brane. We indicate collectively by $H$, $\psi_\alpha$, and $A_\mu$ the scalars, 
fermions and vectors, and by
$m$ the mass parameters (any mass parameter, including the Higgs mass). According to the ansatz of 
eq.~(\ref{fluctuations}),
and keeping the  radion fixed for simplicity,
the induced metric at the TeV brane is 
\beq
{g_\pi}_{\mu\nu}=e^{-2\sigma(\pi)} \bg_{\mu\nu}(x).
\eeq
so that the low energy effective action including the $y=\pi$ boundary contribution is
\beq
{\cal L}^{eff}=\Bigl \{\frac{2M_5^3}{k}\left (1-e^{-2\sigma(\pi)}\right )\sqrt{\bg}\R(\bg)+\sqrt{g_\pi}{\cal L}_\pi({g_\pi}_{\mu\nu},
H,
\psi_\alpha,\A_\mu,m)\Bigr \}.
\label{warp}
\eeq
Notice that the metric that couples to matter is rescaled with respect to the one that appears in the Einstein term,
as if different length units were used in the two actions. To make
physics manifest it is useful to perform a constant Weyl transformation on the matter action. This is just
a  reparametrization of field variables corresponding to a change of the length unit. Indicating by $w$ the Weyl 
rescaling
parameter, in 4 dimension we have that the various fields transform as
\beq
\left ( g_{\mu\nu}, \, H, \, \psi_\alpha, \, A_\mu \,\right ) =\left ( w^{-2}g'_{\mu\nu}, \, w H', \, 
w^{3/2}\psi'_\alpha, \, A'_\mu \,
\right ).
\eeq
Indicating collectively the fields with $\Phi$ and the rescaled ones with $\Phi'$ a generally covariant action $S$ satisfies the
relation
\beq
S(\Phi, m) = S(\Phi',\frac{m}{w}).
\label{weyl}
\eeq
One can easily check this result by considering the free scalar Lagrangian 
\beq
\sqrt{g}\left ( g^{\mu\nu}\partial_\mu H\partial_\nu H-m^2 H^2\right )
\eeq
In the case of the matter contribution in eq. (\ref{warp}) we make a rescaling with $w=e^{\sigma(\pi)}$ such that 
${g_\pi}_{\mu\nu}=w^{-2}\bg_{\mu\nu}$,
$H=w H'$, \dots and write it as
\beq
\sqrt{\bg}{\cal L}_\pi(\bg_{\mu\nu},H',\psi'_\alpha,A'_\mu, m e^{-\sigma(\pi)}).
\eeq
In these new variables it is evident that all mass parameters in the matter Lagrangian are redshifted with respect 
to the Planck
mass $M_4\simeq \sqrt{M_5^3/k}$. Now the importance of the RS mechanism has become very concrete. We stress once 
again that
the basic reason for this result is that the 4-dimensional Planck mass is dominated by a contribution from the 
region
of low red shift, while the SM lives deep inside a region of high red shift. As a matter of fact the metric $\bg_{\mu\nu}$
that appears in the 4D Einstein term coincides with the induced metric on the Planck brane ${g_0}_{\mu\nu}$. We will 
later explain that  all these facts corresponds to the localization of the 4D graviton near the Planck brane.

We want to emphasize that eq.~(\ref{weyl}) implies that only ratios of mass parameters are observable in a theory
of gravity, since their overall normalization can be trivially changed by a field redefinition. This is the same 
situation
we have in the SM, where all CP violating phases, but one, transform under field redefinitions and can thus 
be eliminated.  Only the ratio of Planck and weak scales, the hierarchy,
is an obervable. Then we can have an alternative view point of our result, where the original scale in the Lagrangian 
is the weak scale and the Planck mass is a blue shifted derived scale.
Consider indeed a rescaling with $w=e^{\sigma(\pi)}$ on our original Lagrangian and on our original solution. Now 
all mass 
parameters in
the starting Lagrangian are of order TeV; for instance $M_5\to M_5'=M_5/w$. Moreover the warp factor is 1 at $y=\pi$ 
and $e^{\sigma(\pi)}$
at $y=0$. The 4D Planck mass is now written as
\beq
M_4^2=\frac{(M_5')^3}{k'}\left (e^{2k'r_c'\pi}-1\right ).
\label{primed}
\eeq
This picture makes more evident that the the hierarchy is generated  by a ``dilution'' mechanism, like in the ADD 
scenario. 
In eq.~(\ref{primed}), $M_4^2$ comes out large because of the exponential growth of the ``volume'' towards the 
Planck brane.


After the derivation of the effective Lagrangian we can better understand the meaning of the tunings imposed on the
boundary tensions. As we will show below, the KK spectrum is quantized in units of the radion  
VEV $\langle \mu \rangle$. For $E\ll \langle \mu \rangle$ the system can be described by the zero modes $\bg_{\mu\nu}$ and $\mu$.
Now, if we add to the brane tensions a perturbation which is {\it small}  enough not to excite the KK modes we 
should be able
to accurately describe its effects just in terms of the zero modes. Consider then the following perturbations of the tensions
parametrized by $\alpha,\beta \ll 1$
\beq
\tau_0 =24M_5^3k\left (1+\alpha\right )\quad\quad \tau_\pi=-24M_5^3k\left (1-\beta\right ).
\eeq
At lowest order, the correction to the effective Lagrangian is simply obtained by substituting 
eq. (\ref{fluctuations}) 
in the terms proportional
to $\alpha$ and $\beta$ in the the original Lagrangian
\beq
\Delta {\cal L}^{eff}_4=-24\left (\frac{M_5}{k}\right )^3\sqrt{g}\left (\alpha k^4+\beta\mu^4\right ) .
\label{correctedpotential}
\eeq
By integrating out the  massive KK there arise extra corrections that have either derivatives in them or are of 
higher order in $\alpha,\beta$.
The above equation represents the leading correction to the potential. It is evident that the two tunings of the 
brane tensions, $\alpha=\beta=0$, correspond to
\begin{enumerate}
\item vanishing radion potential
\item vanishing 4-dimensional cosmological constant.
\end{enumerate}
Now, the second requirement is truly necessary since the the cosmological constant  is experimentally extremely small
 $\Lambda_4^4\lsim (10^{-3} {\rm eV})^4$.  On the other hand, by the first requirement the radion is a massless 
Brans-Dicke field
leading to a new  unacceptable long range force. The first tuning is not only useless but experimentally
ruled out. The basic RS model must be modified in order to give a mass to the radion while retaining the possibility
to fine tune the 4D cosmological contant to zero. In the modified theory there would remain just one tuning.

One can ask if a modifcation of the brane tensions can lead to a realistic radion stabilization.
Of course we already know from the derivation of the RS solution that with detuned tensions we would not be able
to find a static solution which is also flat from the 4D viewpoint. But it is useful to study this more 
quantitatively.
Notice that $\beta$ gives rise to a simple quartic potential for $\mu$, so it would seem that the only stationary 
point is
$\mu=0$, which is not interesting. The situation is however slightly more subtle since $\mu$ mixes kinetically with 
the graviton.
In order to easily read the dynamics of $\mu$ it is useful to perform a Weyl rescaling of the metric
\beq
\bg_{\mu\nu}\to \frac {\bg_{\mu\nu}}{1-\frac{\mu^2}{k^2}}
\eeq
to go in the Einstein frame in which the gravitational kinetic term is exactly $2(M_5^3/k) \sqrt {\bg}\R(\bg)$ with 
no radion contribution.
At the same time the radion kinetic term receives a negligibly small modification, while the potential term becomes
\beq
\Delta V=-\Delta {\cal L}^{eff}_4=24\left (\frac{M}{k}\right )^3\sqrt{g}\frac{\alpha k^4 + \beta \mu^4}{(1-\frac{\mu^2}
{k^2})^2}.
\eeq
This potential is stationary at $(\mu^2/k^2)=-\alpha/\beta$. One can easily check that this stationary point 
corresponds to a minimum
only for $\alpha<0$, $\beta>0$. Around the stationary point, $V$ is negative so that the 4D metric in turn will be
$AdS4$. It turns out that both the $AdS4$ curvature $k_4$ and the radion mass $m_\mu$ scale in the same way
$k_4^2\sim m_\mu^2\sim -\alpha k^2$. Then this solution does not look even approximately like the real world, and
a more realistic mechanism of radius stabilization is needed. Before concluding this section
we would like to notice that our simple discussion captures and explains in a physically intuitive way
the results of refs.\cite{kaloper,chacko}, where the full 5D equations of motion in the presence
of detuned brane tensions were studied. 
Our approach also clarifies the results of early studies of RS cosmology \cite{Csaki:1999jh,Cline:1999ts}, 
where a puzzling tuning $\rho_0=-\rho_\pi (\mu/k)^2$ between the energy momentum
densities at the two different branes was found to be needed. In our particular set up we have $\rho_0\equiv 
k^4\alpha$ and
$\rho_\pi\equiv k^4 \beta$, showing that the tuning is just the radion stationarity condition in the absence of 
extra contributions to the potential \cite{Csaki:1999mp}.

\subsection{Radius stabilization: Goldberger-Wise mechanism}
\label{GW}
As we have already remarked, in the RS model  the hierarchy between mass scales at the two boundaries 
depends exponentially on the proper distance $\pi r_c$ between them. A moderate separation
$\pi r_c/L\sim 40$ between the proper length and the $AdS$ length of the 5th dimension is then enough to obtain a 
huge
hierarchy. This would seem a natural way to explain the hierarchy. The implicit assumption behind this conclusion
is  that $r_c$ is the natural dynamical variable to describe the stabilization of the 5th dimension, {\it i.e.} the 
radion potential
is practically a polynomial in $r_c$.
 However we have so far no solid reason to 
believe this is the case. As a matter of fact, in the simple example at the end of the previous section, the natural variable was
the warp factor $\mu/k$ itself. Now,  instead of the proper distance, an observer at the Planck brane
could decide to define her/his distance to the TeV brane through the time $T$ a light signal (or a graviton!)
takes to travel to the TeV brane and back. The result would be
\beq
T=2\int_0^\pi\frac{d y}{\sqrt{g_{00}}}=\frac{2}{k}\left (e^{kr_c\pi}-1\right )\sim \frac{2}{\mu}
\eeq
which is exponentially large as a consequence of time dilation near the TeV brane. According to this perfectly 
acceptable definition, the size of
the extra dimension is controlled by the weak scale $\mu$ itself. Notice  that  the potential of the previous 
section is polynomial in $\mu$.
Such hugely different notions for the size of the 5th dimension arise because the global features of the RS space 
are controlled by curvature. 
We could  have started with a coordinate system in which, according to the above definition of distance, the 5th 
coordinate is parametrized by
\beq
z=  \frac{e^{-kr_c y}}{k}.
\eeq
The metric would then have been
\beq
ds^2=\frac{L^2}{z^2}\left (\eta_{\mu\nu}d x^\mu d x^\nu+ dz^2\right ).
\label{confcoord}
\eeq
where $L=1/k$, the $AdS$ radius, represents the $AdS$ curvature length. In these ``conformal'' coordinates
there is no exponential factor, but the locations of the Planck and TeV branes, respectively $z_0$ and $z_1$, are 
very far apart
\beq
z_0= L \ll L e^{k r_c \pi}=z_1
\eeq
In these new coordinates the radion field is simply associated to the position of the second brane: 
$\mu(x) =1/z_1(x)$.

To solve the hierarchy problem we must then find a mechanims that stabilizes the second brane at 
$z_1\sim 1/{\rm TeV}\gg z_0\sim 1/M_P$.
In order to do so, Goldberger and Wise (GW) \cite{gw} have proposed a simple model involving a 5D scalar field 
$\phi$ with action
\bea
S_\phi&=&\int d^4 xdz
\left \{\sqrt {g}\bigl [-(\partial\phi)^2-m^2\phi^2\bigl ]+ \delta(z-z_0)
\sqrt{g_0}{\cal L}_0(\phi)\right.\\\nonumber
&+&\left.\delta(z-z_1)\sqrt{g_1}{\cal L}_1(\phi)
\right \}
\label{phiaction}
\eea
It is assumed  that
the dynamics of the boundary terms ${\cal L}_{0,1}$ is such as to fix the values 
$\phi(z_0)=\tilde v_0$ and $\phi(z_1)=\tilde v_1$. For instance,  one can take 
${\cal L}_{0,1}=-\lambda_{0,1}(\phi-v_{0,1})^2$ with  $\lambda_{0,1}\to \infty$. The assumption of an infinitely steep 
boundary potential 
simplifies the computations but is not essential \cite{gw,Rattazzi:2000hs}.
In the vacuum, the
field $\phi$ will have a non-trivial bulk profile satisfying
 the 5D Klein-Gordon equation and interpolating between the two boundary values $v_0$ and $v_1$.
The energy associated to this profile  depends on the distance between the two branes, corresponding to a 
non-vanishing
radion potential.
Now, solving the coupled equations of motion of gravity plus $\phi$ exactly is in general difficult. To make
our task easier we can make the simplifying assumption that $\phi$ only induces a small perturbation of the locally
$AdS5$ Randall-Sundrum metric. Quantitatively this is equivalent to requiring the scalar energy momentum  to
be a small perturbation of the 5D cosmological constant
\beq
T^{\phi}_{MN}\sim (\partial _z\phi)^2+m^2\phi^2\sim {\rm max}(v_0^2,v_1^2)\times {\rm max}(k^2,m^2)\ll 
T_{MN}^{vacuum}\sim M_5^3 k^2.
\eeq
Since, in order to generate a big hierarchy we will need $m^2\lsim k^2$, the above relation simply implies
\beq
v_{0,1}^2\ll M_5^3
\eeq
(notice that a scalar field in 5D has mass dimension $3/2$). When this condition holds, $\phi$ is determined at 
leading order
by solving the $\phi$ equations of motion over the unperturbed RS background 
with  Dirichlet  boundary conditions $\phi(z_0)=v_0$ and $\phi(z_1)=v_1$. Because of this second constraint, there 
are no
massless zero modes in the KK tower of $\phi$. As long as the $\phi$ profile is a small perturbation of RS the only 
light states are those of the unperturbed
model: the 4D graviton and the radion. In this respect the procedure of solving the $\phi$ equations of motion
 and calculating the action
on the solution corresponds to integrating out the tower of massive $\phi$ KK modes to obtain an effective Lagrangian
for the light modes $\bg$ and $\mu$. This is what we will now do. For a field configuration that does not depend
on the 4D coordinates the 5D Klein-Gordon equation becomes
\beq
\frac{z^5}{L^2}\partial_z\frac{1}{z^3}\partial_z \phi=m^2\phi
\eeq
and the most general solution is
\beq
\phi=Az^{4+\epsilon}+Bz^{-\epsilon}\quad\quad\quad \epsilon= \sqrt{4+m^2L^2}-2\simeq\frac{m^2L^2}{4} \,.
\label{phisolution}
\eeq
where we are emphasizing that we will be momentarily interested in the limit $\epsilon\sim m^2L^2/4\ll 1$.
The boundary conditions fix
\beq
A=z_0^\epsi\frac{ v_0- v_1 (z_0/z_1)^{4+\epsi}}{1-
(z_0/z_1)^{4+2\epsi}}\quad\quad
B=z_1^{-4-\epsi}\frac{ v_1- v_0 (z_0/z_1)^{\epsi}}{1-
(z_0/z_1)^{4+2\epsi}}
\label{aandb}
\eeq
and eq.~(\ref{phiaction}) evaluated on the solution yields an effective 
potential for the radion $\mu$
\bea
V(\mu)
&=&\frac{1}{1-(\mu L)^{4+\epsi}}\left[{(4+\epsi)\mu^{4}\bigl (\bar v_1-\bar v_0
(\mu L)^\epsi\bigr )^2}\right.\\\nonumber
&+&\left.{\epsi L^{-4}\bigl (\bar v_0-\bar v_1(\mu L)^{4+\epsi}\bigr )^2}\right]\\ 
&=&L^{-4} F(\mu L)\nonumber
\label{vbare}
\eea
where we have made the substitutions $z_0=L$, $z_1=1/\mu$ and
where $\bar v_{0,1}=L^{3/2} v_{0,1}$ are the boundary VEVs in units of the $AdS$ curvature. 
For $\mu L\ll 1$ the potential becomes
\beq
V=\epsi \bar v_0^2L^{-4} +\left [(4+2\epsi)\mu^4 \bigl(\bar v_1-\bar v_0(\mu/\mu_0)^\epsi
\bigr )^2-\epsi \bar v_1^2\mu^4\right ] +{\cal O}(\mu^8 L^4)
\label{vren}
\eeq
which  for $\epsilon >0$ is minimized at
\beq
\mu L \simeq \left (\frac{\bar v_1}{\bar v_0}\right )^\frac{1}{\epsilon}.
\label{muvev}
\eeq
The hierarchy $\langle \mu\rangle L\sim M_W/M_P= 10^{-17}$
can be naturally obtained for fundamental parameters
not much smaller than one (ex. $\bar v_1/\bar v_0\sim 1/10$ and $\epsi\sim 1/20$).
The hierarchy naturally arises because of the non-analytic dependence of eq.~(\ref{muvev}) on $\epsi$ .
Some comments are now in order.
\begin{enumerate}
\item Notice that the relevant part of the potential (the term in square brackets in
eq.~(\ref{vren})) has the form
$\mu^4 P(\mu^\epsi)$. This is basically a 
quartic potential modulated by a slow evolution of the effective coupling $\lambda\equiv P(\mu^\epsi)$.
Notice indeed that for $\epsi\ll 1$ one has $\mu^\epsi \sim 1+\epsi \ln \mu +\dots$, so that the dependence on $\mu$ is reminiscent of a slow RG evolution.  Stability of the potential at large $\mu$ corresponds to $\lim_{\mu\to \infty}P(\mu)>0$.
Because of the slow evolution of $P$, a minimum of the potential will exist very close to the point where $P$ crosses zero to become negative, see 
eqs. (\ref{vren},\ref{muvev}).
These properties are in full analogy with the Coleman-Weinberg (CW) \cite{coleman} mechanism of dimensional transmutation. 
There, quantum corrections
to the effective potential cause the quartic coupling to turn negative at some scale. Since this happens through the slow
logarithmic RG evolution, broadly different mass scales can arise, making the CW mechanism very interesting to explain the weak to Planck
scale hierarchy. It seems that the GW mechanism works qualitatively in the same way.
\item Expanding at second order around the minimum and using the unperturbed kinetic Lagrangian we find the radion mass
\beq
m_\mu^2\simeq \frac{2}{3}{\bar v_1}^2 \epsi^{3/2} \langle \mu \rangle ^2\ll \langle \mu \rangle ^2
\eeq
indicating that in the model at hand the radion is much lighter than the other KK resonances and potentially more
interesting for collider phenomenology.
\item The potential at the minimum is dominated by the fist term in eq.~(\ref{vren}): $V_{min}\simeq \epsi {\bar v_0}^2 L^{-4}\sim M_4^4$, far too large!
However we can now go back and slightly modify the brane tensions as we did in the previous section. For small $\alpha$ and $\beta$, at leading order  this amounts to adding the contribution in eq.~(\ref{correctedpotential}) to eq.~(\ref{vren}). 
Around the minimum of the GW potential $\mu$ is massive so that 
$\beta\not = 0$ causes only a small shift  in $\langle \mu \rangle$. We no longer need to tune $\beta=0$ after the radius is 
stabilized. Moreover both $\alpha$ and $\beta$ cause a shift in the potential at the minimum. 
By properly choosing one  combination  of them  we can cancel 
the potential at the minimum.  This is of course a fine tuning, but it is just the cosmological constant problem, which we never promised to solve. 
After radius stabilization the only fine tuning left in the model is the one associated to the vanishing cosmological constant.
\item The minimal GW potential gives rise to a stable minimum only for $\epsi \simeq m^2L/4>0$. However one could also obtain stable minima for $\epsi<0$
for a proper range of the detuning parameter $\beta$.
\end{enumerate}

Concerning the last remark  one may worry that $\epsilon\sim m^2<0$ will lead to instabilities in the bulk. If  $|m^2|$ is small enough, however, no instability arises.
Concerning this result it is instructive to consider the following exercises.
\begin{itemize}
\item Show that no instability is generated for $m^2>-4k^2+({\rm small})$ 
(hint: study the KK spectrum for the excitation $\Delta \phi$ around the GW background
remembering that $\Delta \phi=0$ at the boundaries).
\item Study the same problem with a flat 5th dimension.
\end{itemize}

\subsection{Kaluza-Klein spectrum}
In this section we shall study the KK spectrum of the gravitational field.
As for the flat case we studied earlier,  it is convenient to use the 5D gauge freedom
to eliminate as many degrees of freedom  from  $g_{\mu 5}$ and $g_{55}$ as possible.
It is then easy to check (for instance by working at the linearized level) that
without loss of generality we can use the following parametrization 
\beq
ds^2=e^{-2k|y|r_c- 2f(x)e^{2k|y|r_c}}\left (\hat g_{\mu\nu}(x,y)d x^\mu d x^\nu\right )+ 
\left (1+2f(x)e^{2k|y|r_c}\right )^2dy^2
\label{fg}
\eeq
Because of the orbifold projection we can consistently set $g_{\mu 5}\equiv 0$.
On the other hand for $g_{55}$ we can eliminate all the modes but one, the radion.
This one mode remaining in $g_{55}$ can however be parametrized at our convenience, and the choice in 
eq.~(\ref{fg}) as an advantage that we will now explain.
Since we are mostly interested in the $J=2$ modes, consider first substituting eq.~(\ref{fg}) with $f(x)=0$
in the RS action. After changing the 5th coordinate to $z=Le^{-k|y| r_c}$, one finds
\beq
S(g)=4M_5^3L^3\int_L^{z_1}\sqrt{\hat g}\frac{dz}{z^3}\Bigl [\R_4(\hat g)+\frac{1}{4}\left ((\partial_z \hat g^{\mu\nu})
(\partial_z \hat g_{\mu\nu})+(\hat g^{\mu\nu}\partial_z \hat g_{\mu\nu})^2\right )\Bigr ]\,.
\label{pfrs}
\eeq
Notice that all terms with no derivative acting on $\hat g_{\mu\nu}(x,y(z))$  have judiciously 
cancelled out when expanding around the RS solution.
We already know that it should have been this way: any constant $\hat g_{\mu\nu}$ solves the equations
of motion, so that the action for $\hat g_{\mu\nu}$ must involve only gradient terms.
When $f\not =0$ we find an extra contribution to be added  to eq.~(\ref{pfrs}) and starting at quadratic order in $f$
\beq
\Delta S=-12 \frac{M_5^3}{L} \left (z_1^2 -L^2\right ) \partial_\mu f \partial ^\mu f +O(f^3,f^2 h_{\mu\nu},\dots)
\label{kinf}
\eeq
where $h_{\mu\nu}=\hat g_{\mu\nu}-\eta_{\mu\nu}$. Notice that there is no kinetic mixing between $f$ and the metric $h_{\mu\nu}$.
In other words, the graviton is automatically in the Einstein frame. 
This is because of our specific parametrization of the scalar mode in eq.~(\ref{fg}). This should be contrasted to the parametrization
in eq.~(\ref{fluctuations}) which lead to a small $O(\mu^2 L^2)$ mixing between radion and graviton.
Anyway, neglecting  $O(\mu L)$, terms and with the identification (cfr. eqs.~(\ref{fluctuations},\ref{radion})) 
\beq
\mu(x)= k e^{-k\pi r_c- f(x)e^{2k\pi r_c}}=ke^{-\sigma(\pi,x)}
\eeq
eq.~(\ref{kinf}) reproduces the radion kinetic term of eq.~(\ref{effectiveLradion}).

Let us now focus on the $J=2$ modes. Expanding eq.~(\ref{pfrs}) at quadratic order in $h_{\mu\nu}=
\hat g_{\mu\nu}-\eta_{\mu\nu}$ we find 
\beq
M_5^3L^3\int_L^{z_1}\frac{dz }{z^3}\Bigl ( h_{\mu\nu}K^{\mu\nu\rho\sigma}h_{\rho\sigma}-h^{\mu\nu}\partial_z^2 h_{\mu\nu}+h_\mu^\mu \partial_z^2 h_\nu^\nu\Bigr )
\label{linearizedRS}
\eeq
where $K^{\mu\nu\rho\sigma}$ is the 4-dimensional $J=2$ kinetic operator shown in the first
line of eq.~(\ref{linearizedKK}). Notice that the mass operator has the Fierz-Pauli form.
From eq.~(\ref{linearizedRS}) the equation for the eigenmodes $\psi_n(z)$ is then simply
\beq
-z^3\partial_z \frac{1}{z^3}\partial \psi_n(z)= m_n^2 \psi_n(z)
\label{eigenRS}
\eeq
with boundary conditions
\beq
\partial_z\psi_n\vert_{z=L}=\partial_z\psi_n\vert_{z=z_1}=0,
\eeq
as $h_{\mu\nu}$ is even under the orbifold parity and   its action does not contain boundary terms.

For $m_n\not = 0$ the general solution to eq.~(\ref{eigenRS}) is written in terms of Bessel functions
\beq
\psi_n(z)=\frac{z^2}{L^2}\left [ J_2(m_n z)+b_n Y_2(z m_n)\right ]
\label{bessel}
\eeq
and the boundary conditions enforced by using the identity
\beq
\partial_z \psi_n\propto z^2 \left [ J_1(m_n z)+b_n Y_1(z m_n)\right ].
\eeq
We then find 
\beq
b_n=\frac{J_1(m_n L)}{Y_1(m_n L)}
\label{bn}
\eeq
while the eigenvalue equation is simply
\beq
J_1(m_n L) Y_1(m_n z_1)-Y_1(m_n L) J_1(m_n z_1)=0.
\label{exacteigen}
\eeq
In order to focus on the phenomenologically interesting ``light'' modes, satisfying $m_n L\ll 1$, we need 
the asymptotic behaviour of the Bessel functions
\bea
x\ll 1 \quad\quad \rightarrow\quad\quad \left \{\begin{array}{l} J_2(x)=\frac{x^2}{8}+O(x^4)\\
                                                 Y_2(x)=-\frac{4}{\pi x^2}-\frac{1}{\pi}+O(x^2)\end{array}\right.\cr
x\gg 1\quad\quad \rightarrow\quad\quad \left\{ \begin{array}{l}J_2(x)=\sqrt{\frac{2}{\pi x}}\cos(x-\frac{5}{4}\pi)\cr
                                                 Y_2(x)=-\sqrt{\frac{2}{\pi x}}\sin(x-\frac{5}{4}\pi)\end{array}\right. .\cr
\label{asymptotic}
\eea
In the limit $m_n L\ll 1$ eqs.~(\ref{bn}) and (\ref{exacteigen}) reduce respectively to
\beq
b_n\simeq (m_n L)^2\pi/4\ll 1
\eeq
and
\beq
J_1(m_n z_1)= 0
\eeq
The solutions to the last equation are quantized in units of $1/z_1\equiv\langle \mu \rangle$
\beq
m_n=c_n\frac{1}{z_1}\simeq (n+\frac{1}{4})\frac{\pi}{z_1},
\eeq
where the last identity is valid asymptotically for $n\gg 1$ (cfr.  eq.~(\ref{asymptotic})) but works very well already for $n=1$: 
$c_1\simeq 1.21 \pi$. Notice that because $b_n\ll 1$, $Y_2$ makes a negligible contribution to $\psi_n$ in the region
$m_n z\gg 1$ where $\psi_n$ oscillates. In the region $zm_n\ll 1$, the $Y_2$ contribution is relatively important (in fact dominant),
but this region contributes negligibly to the normalization of the eigenfunctions.

Indeed, by eq.~(\ref{linearizedRS}) the norm of the modes is  
\bea
||\psi_n||^2
&=& \int_{z_0=L}^{z_1}\frac{2 z dz}{L^2} \left [J_2(m_n z)+b_n Y_2(m_nz)\right ]^2\simeq
\int_0^{z_1}\frac{z dz}{L^2}\left [J_2(m_n z)\right ]^2\cr
&=& \left (\frac{z_1}{L}\right )^2 J_2(m_n z_1)^2.
\eea
Neglecting contributions of relative size $O(L^2/z_1^2)\sim m_{weak}^2/M_P^2$, we have approximated 
$||\psi_n||$ by its value in the limit $z_0=0$, in which the Planck brane  has an infinite proper distance from the TeV brane
(and an infinite relative blueshift, see eq.~(\ref{confcoord})). $||\psi_n||$ is dominated by the region $1/m_n < z< z_1$, where
the Bessel functions have an oscillatory behaviour. A mode with mass $m_n$ is not very sensitive to the region  $z< 1/m_n$.
In particular the  modes remain normalizable and the spectrum discrete even for $z_0\to 0$. 
Notice finally that the orthonormal modes ${\hat \psi}_n(z)\equiv \psi_n(z)/||\psi_n||$ satisfy ${\hat \psi_n}(z_1)= 
{sgn(J_2(m_n z_1))} z_1/L$,
so that all modes couple with equal strength to the fields on the SM brane. 
 
The massless mode, see eq.~(\ref{eigenRS}), simply corresponds to a constant $\psi_0=1$
with norm
\beq
||\psi_0||=L^2\int_{z_0=L}^{z_1}\frac{2 dz}{z^3}=\left (\frac{1}{L^2}-\frac{1}{z_1^2}\right )
\label{psi0}
\eeq
which is dominated at small $z$. Unlike the massive modes, $\psi_0$ is localized at the Planck brane: $||\psi_0||$ diverges for $z_0\to 0$ but remains finite
for $z_1\to \infty$. From eq.~(\ref{pfrs}) we have that the 4D Planck mass is  $M_4^2=M_5^3L^3 ||\psi_0||^2=M_5^3L[1-(L/z_1)^2]$, which is the result we obtained 
previously. Notice also that for $z_0\to 0 $ the 4D Planck mass diverges, corresponding to the zero mode 
graviton being localized infinitely far away from the TeV brane.

Using the above results, but working with
the canonically normalized KK gravitons ${\bar h}_{\mu\nu}^{(n)}$ and radion $\bar f$ the coupling to the TeV brane energy momentum tensor is written as
\beq
{\cal L}_{int}= -\frac{1}{2}\left \{\frac{{\bar h}_{\mu\nu}^{(0)}}{M_4}+\sum_{n\geq 1}\frac{{\bar h}_{\mu\nu}^{(n)}}{\Lambda}\right \} T^{\mu\nu}+\frac{\bar f}{\sqrt{24} \Lambda}
T_\mu^\mu
\label{KKcoupling}
\eeq
where $\Lambda\equiv (M_5 L)^{3/2}/z_1$. Eq.~(\ref{KKcoupling}) is the basic equation to study the collider implications of the
 RS model \cite{Davoudiasl:1999jd}.
The interactions and spectrum of the $J=2$ modes are fully described by two parameters, $\Lambda$ and $\langle \mu \rangle=1/z_1$. For radion
phenomenology two extra parameters are needed, one is the radion mass, which depends on the stabilization mechanism, the other
is a radion-Higgs mixing parameter $\xi$ \cite{grw2}. Basically the parameter $\xi$ accounts for the fact that the energy momentum
tensor for a scalar $H$ is defined up to an ``improvement term'' $\Delta T_{\mu\nu}=\xi (\partial_\mu\partial_\nu-\eta_{\mu\nu}\partial^2)H^2$.
A non zero $\xi$ induces a kinetic mixing between radion and Higgs after electroweak symmetry breaking.

A little exercise one can do with the interaction Lagrangian in eq.~(\ref{KKcoupling}) concerns the validity of perturbation theory.
A simple quantity to calculate (estimate) is the decay width into SM particles (living at the TeV boundary). For the $n$-th mode we find
\beq
\Gamma_{(n)}\sim \frac{m_n^3}{8\pi \Lambda^2}.
\eeq
Notice that $\Gamma_{(n)}$ grows with $n$, so that for $n$ large enough the nearby resonances will start overlapping.
When this takes place, by definition,  perturbation theory breaks down:  quantum corrections to the spectrum (the widths)
make the very concept of individual KK resonances useless. Using the above equations we find that $\Gamma_{(n)}<m_{n+1}-m_{n}$
is satisfied for masses below
\beq
m_n< M_5 \frac{L}{z_1}.
\eeq
This shows that the UV cut-off for physics at the TeV scale is just the redshifted Planck mass. Of course by starting 
off in the primed Weyl frame, discussed above eq.~(\ref{primed}), this is the obvious result.

\subsection{Strong coupling puzzle}
\label{strong}
At the end of the previous section we have shown by a simple argument that the RS model becomes strongly
coupled at a fairly low scale ${M'}_5= M_5L/z_1$. If we we want to explain the hierarchy by the ratio $L/z_1$,
then ${M'}_5\sim $TeV very much like in the ADD scenario. Although such a low-cut off limits predictivity,
as long as  $ M_5'$ is somewhat bigger than the mass of the lightest KK, some
control is retained: roughly a number of modes $\sim { M'}_5 z_1= M_5L$ remains weakly coupled.
From the conceptual viewpoint, however, the presence of this low cut-off can be confusing, when not properly interpreted.
One basic puzzle is that ${M'}_5$ depends on the location of the TeV brane. Moreover as $z_1\to \infty$
and half of $AdS5$ is recovered, ${ M'}_5$ goes to zero, as if there was no energy range where gravity on $AdS$ makes
sense as an effective field theory. The origin of this  puzzle is  that we are working on a 
curved space
where particle propagation involves large or possibly infinite momentum blue-shift. As we will explain, since the puzzle 
arises when considering
the global aspect of our spacetime, a proper resolution cannot do without accounting for
the locality of the interactions.
 
It is instructive to go back and consider the motion of a particle in $AdS$ space. The equations for the geodetic in conformal coordinates are
 \beq
 {\ddot x}^\mu-2\frac{{\dot x}^\mu \dot z}{z}=0 \quad\quad\quad {\ddot z} -\frac{{\dot z}^2}{z}+\frac{{\dot x}^\mu 
 {\dot x}^\nu \eta_{\mu\nu}}{z}=0
 \label{geodetic}
 \eeq
 where by the dot we indicate the derivative with respect to the affine parameter $\lambda$. Let us focus on the massless case.
 As $\lambda$ is arbitrary, we can choose it such that $d x^M/d\lambda$ coincides with the 5-momentum $P^M$.
By this choice we have, in particular, $d/d\lambda = P^5 d/d z$ and by using the zero mass condition $P^M P^Ng_{MN}=0$ we can write eq.~(\ref{geodetic}) and its solution as 
\beq
z \frac{d P^M}{dz}=2 P^M \quad\quad \rightarrow\quad\quad P^M=\left (\frac{z^2}{L^2}\right ) {\bar P}^M.
\eeq
 This is the momentum in the $AdS$ coordinates, but a more physical quantity is the momentum $P^M_{phys}$ seen in the frame of a 
free falling observer at rest at  $z$: $P_{phys}^M= P^M(L/z)=  \bar P^M(z/L)$.
Viewing the particle as a wave, the increase in momentum is simply due to the homogeneous contraction $L/z$ of
all lengths when moving down the $AdS$ throat, see Fig.~(\ref{blueshift}). Moving to larger $z$ is like going backward in time in 
Friedmann-Robertson-Walker  (FRW) cosmology. This analogy is not accidental as both $AdS$ and FRW 
have conformally flat metrics. Now,  consider two particles starting with momenta  $\bar P_1^M$ and $\bar P_2^M$ at the 
Planck brane. If they collide 
after falling at a bulk position $z$, the center of mass energy of the collision will be
$s(z)=-2P_1^M P_2^Ng_{MN}=-2 \bar P_1^M \bar P_2^N\eta_{MN} (z/L)^2= s(0)(z/L)^2$. Then by starting by a sub-Planckian energy 
${\sqrt s(0)}<M_5$ we can produce
super-Planckian collisions if $z$ is large enough. Of course this also means that a Planck brane observer must wait 
 a relatively long time $T>z$ in order to oberve this collision. In a moment we will see that  this time delay 
is
the central point to discuss ``strong coupling'' at the quantum level.
Notice also that in the case of the compact  RS model $z$ is bounded by $z_1$, so that for ${\sqrt {s(0)}}<M_5 
L/z_1$ the Planck mass is never
exceeded in the bulk. Once again we have found that the critical energy corresponds to the redshifted Planck mass 
$M_5'$.

Consider now the limit $z_1\to \infty$. At $z=\infty$ the metric of 4D slices vanishes, similarly to what happens to $g_{00}$ 
in the Schwarzschild
solution in GR. Also, at the point $z=\infty$ there is  a horizon, the $AdS$ horizon. In fact  a particle falling 
from the Planck brane takes an infinite
Planck brane time to reach $z=\infty$, but the proper time experienced by the particle is finite $\tau = \pi L/2$. 
This is completely analogous to the
Schwarzschild case.
The model so defined is named RSII \cite{rs2}, so that the model with two branes we have considered so far is named 
RSI. 
If we assume that the SM is instead localized at the Planck brane (and thus give
up explaining the hierarchy by redshifts) the RSII model
represents an ``alternative to compactification''. This is because, although the 5th dimension is non-compact,
there is still a normalizable 4-dimensional graviton, see eq.~(\ref{psi0}), dominating the IR behaviour of gravity. 
Eq.~(\ref{psi0}) should be contrasted to
the flat case, in which $||\psi_0||$ grows with the radius $R$, so that the zero mode decouples in the infinite 
volume limit.

\begin{figure}[t]
$$\includegraphics[width=4cm]{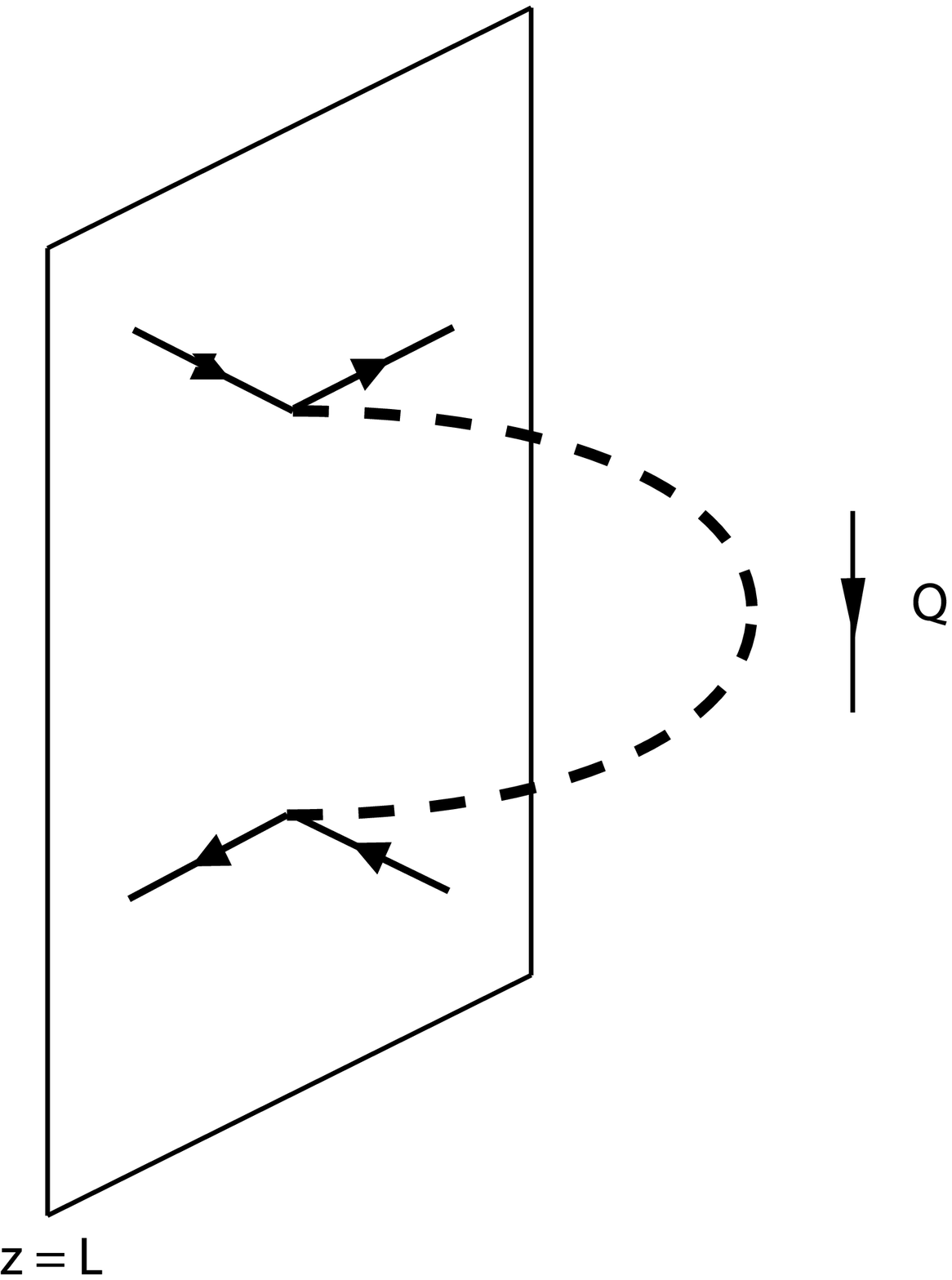}\qquad\qquad\qquad\qquad\qquad\qquad\includegraphics[width=5cm]{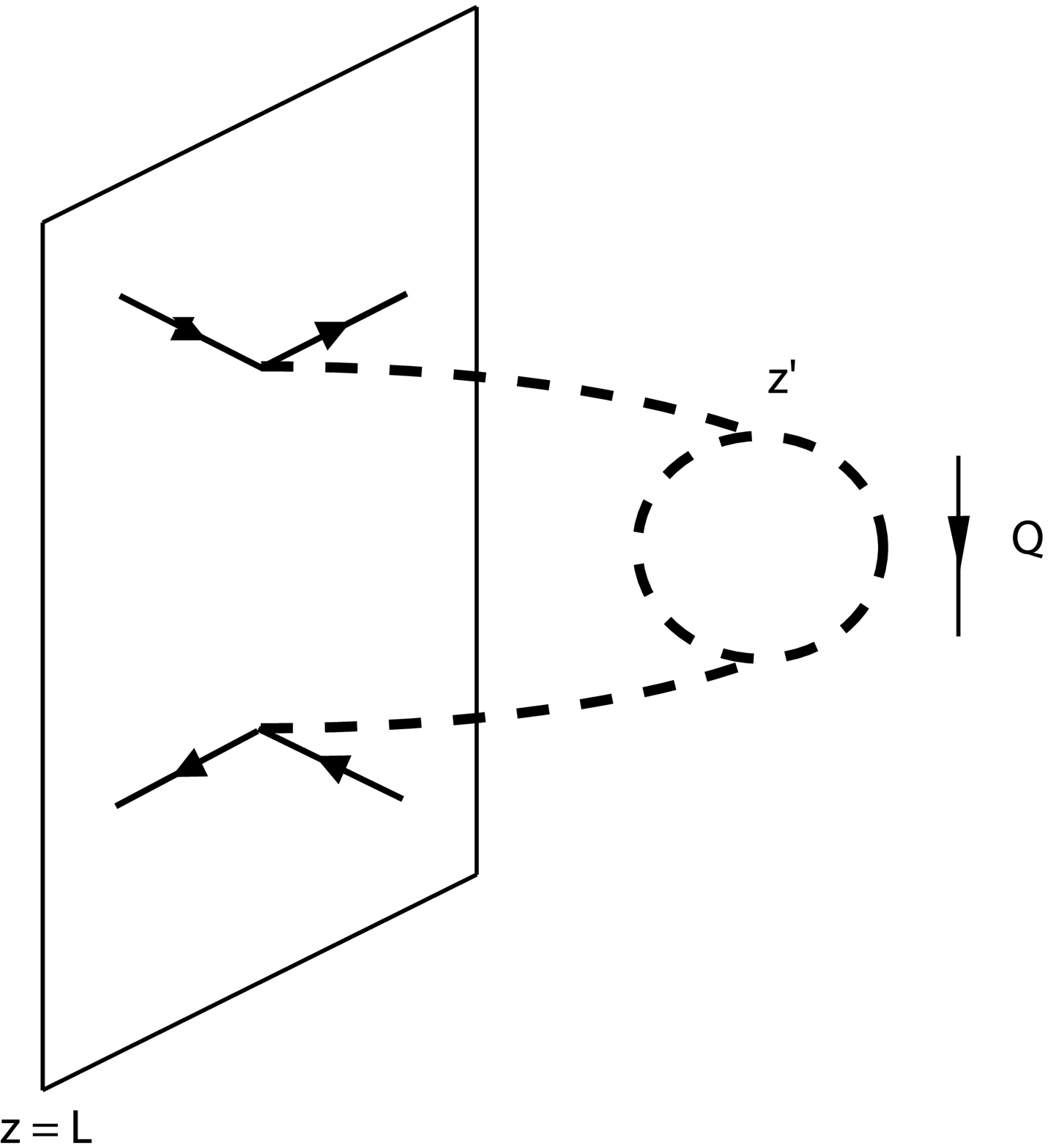}$$
\parbox{7cm}{\caption{\label{tchannel}\em Amplitude for the scattering of fermions localized on the Planck brane 
via $t$-channel
5D graviton exchange.}}\hspace{0.5cm}\hfill
\parbox{7cm}{\caption{\label{bulkloop}\em Leading quantum gravity correction to the process of Fig.~(\ref{tchannel}) .
}}
\end{figure}
The RSII model seems a very interesting way to view 4D gravity.
In this model however a particle falling from the Planck brane can undergo virtually infinite Redshift before 
colliding.
Then if we blindly applied the above definition of UV cut-off we would conclude that RSII
is not a viable effective field theory! And we would be wrong. The point is that the notion of ``which energy is 
Planckian'' must be a local one.
As we have already stated, a Planck brane observer, while working with initial states that have ${\sqrt s}\ll M_5$,
must set up an experiment that probes deep into the 5D bulk in  order to see quantum
gravity effects. An instructive way to understand how things work is to consider
a scattering process induced by graviton exchange among particles localized on the Planck brane. Consider first a $t-$channel process at fixed angle,
for which the exchanged graviton has  virtuality $Q={\sqrt {-t}}\sim {\sqrt s}$. A particle with virtuality $Q$ can exist only for a time $\sim 1/Q$,
so that by causality it probes the bulk no further than $z_Q\sim 1/Q$. The virtual momentum is then only blueshifted to
\beq
Q(z)=Q\frac{z}{L}\lsim Q\frac{z_Q}{L}\sim \frac{1}{L}
\eeq
so that perturbativity of the process is maintained as long as $1/L\ll M_5$. But this was our original requirement in order to 
trust
the RS solution. By the above simple argument we expect the leading gravitational loop corrections to the above process to be
controlled by
$Q(z_Q)/M_5=1/(M_5L)$ independent of the scale of the kinematical parameters $s$ and $t=Q^2$. This result is remarkable: the
leading gravitational corrections are scale invariant. This does not happen by chance and is related to
the possibility to interpret the RSI and RSII models according to the $AdS$/CFT correspondence \cite{maldacena2,apr,Rattazzi:2000hs,Perez-Victoria:2001pa}. 
As shown in Fig.~(\ref{bulkloop}), we can make our arguments slightly more concrete by considering the 1-loop correction to our $t-$channel
process. The crucial remark \cite{apr} is that the brane to bulk propagator (working in momentum space along the brane and in position space 
along the 5th dimension) behaves like
\beq
G(z=z_0, z', Q)\propto e^{-Q (z'-z_0)}
\eeq
while the local loop expansion parameter is $[Q(z')]^3/M_5^3$. Taking the brane to bulk propagator into account, the 1-loop 
correction diagram will be of order
\beq
e^{-Q(z'-L)} \left (\frac{Q z'}{M_5 L}\right ) ^3< \frac{1}{(M_5L)^3}
\eeq
consistent with our previous result. Notice that both the above inequality and the previous argument apply only for $Q < 1/L$. When $Q> 1/L$
curvature is unimportant (the relevant length scale $1/Q$ is short) so that the loop expansion parameter is just
that of flat space $\sim Q^3/M_5^3$. We conclude that a Planck brane observer studying $t$-channel graviton exchange
sees quantum gravity becoming important at the most obvious scale: $M_5$. By the same argument, we would also deduce that
an observer sitting on a probe brane at position $z$ sees gravity becoming strong at a local scale $M_5(z)=M_5 z/L$.

The situation for $s$-channel processes is somewhat different. Here with enough energy resolution one
could produce, even on the Planck brane, individual KK-modes and measure their widths, being thus able to decide whether there is strong dynamics.
The required energy resolution $\Delta E \lsim 1/z_1$, corresponds, by the indetermination principle, to a time scale $T>z_1$
during which  a signal can travel from the Planck to TeV brane.  To be more definite consider the annihilation process $\lambda \bar \lambda\to
\chi\bar \chi$ for fermions living on the Planck brane.
The $s-$channel amplitude is  
\bea
A(s)&=& \frac{1}{4}T^{\mu\nu}({\rm in})T^{\rho\sigma}({\rm out})\langle h_{\mu\nu}h_{\rho\sigma}\rangle\cr
 &=&\frac{1}{4}T^{\mu\nu}({\rm in})T^{\rho\sigma}({\rm out})\cr
&&\sum_{n} \frac{|\hat \psi^{(n)}(z=L)|^2}{M_5^3L} 
\frac{\Pi_{\mu\nu\rho\sigma}{(m_n)}(s)}{s-m_n^2+im_n \Gamma_n}
\label{braneamplitude}
\eea
where $\Pi{(m_n)}$ is the $J=2$ projection operator defined in eq.~(\ref{massiveprop}) and where for simplicity we have 
neglected the radion contribution.
The wave function at the Planck brane
is  $|\hat \psi^{(n)}(z=L)|^2\sim m_n (L^2/z_1)$ for the massive modes, but for the zero mode we have 
$|\hat \psi^{(0)}(z=L)|^2\sim 1 $.
The lighter massive modes, being localized away from the Planck brane, couple much more weakly than the zero mode. In
this respect, the production of these modes is very suppressed. However even with this tiny coupling one could in principle study 
the production of individual resonances and measure their width. In fact we do not want to stress to much the wave function 
suppression, as it is a specific feature of the graviton KK. For instance, in the case of a bulk  vector this  suppression 
is practically absent \cite{Pomarol:1999ad}. The point we want to stress here concerns instead the energy resolution of
 the experiment. In order  to proceed we need the explicit expression of the brane to brane propagator
 as a sum over massive KK , 4D graviton and radion contributions
\beq
M_5^3\langle h_{\mu \nu} h_{\rho\sigma}\rangle =A(s)\Pi_{\mu\nu\rho\sigma}^{m\not = 0} +\frac{z_1^2 }{z_1^2-L^2}\frac{\Pi_{\mu\nu\rho\sigma}^{ m= 0}}{sL}
+\frac{L^2}{z_1^2-L^2}\frac{\eta_{\mu\nu}\eta_{\rho\sigma}}{ 6 s L}
\label{fullampl}
\eeq
where 
\beq
\Pi_{\mu\nu\rho\sigma}^{m\not = 0}=\frac{1}{2}\left (\etamr\etasn+\etams\etarn\right )-\frac{1}{3}\etamn\etars
\eeq
is the massive spin 2 projector of eq.~(\ref{massiveprop}) after eliminating the irrelevant longitudinal parts,
while $\Pi_{\mu\nu\rho\sigma}^{m= 0}$ is the massless projector given 
in eq.~(\ref{masslessprop}). Notice that the radion contribution, the third term, is suppressed by a power $L^2/z_1^2$. 
The massive KK amplitude is 
\bea
A(s)&=&\left \{\frac{1}{2\sqrt s}\frac{Y_1(\sqrt s z_1) J_2(\sqrt s L)- J_1(\sqrt s z_1) Y_2(\sqrt s L)}
{Y_1(\sqrt s z_1) J_1(\sqrt s L)- J_1(\sqrt s z_1) Y_1(\sqrt s L)}-\frac{1}{ s L}\frac{z_1^2}{z_1^2-L^2}\right \}\cr 
&\equiv& \frac{F(s)}{2\sqrt s}-\frac{1}{ s L}\frac{z_1^2}{z_1^2-L^2}. 
\label{totalprop}
\eea
 Notice that $A(s)$, and $F(s)$ have poles on the real positive axis
in correspondence with the massive KK masses. However $A(s)$ does not have a pole at $s=0$, while $F(s)$ does.
For  large $z_1$, on the positive real $s$ axis $A(s)$ is a complicated oscillating function with narrowly spaced poles.
However when $s$ is continued into the complex plane all this structure gives way to the much simpler euclidean behaviour.
By giving $\sqrt s$ a positive finite imaginary part $\sqrt s\equiv q_R+i  q_I$  and by using the asymptotic expansion for
Bessel functions we find
\bea
F(s)&=&\frac{J_2(\sqrt s L)+i Y_2(\sqrt s L)}{J_1(\sqrt s L)+i Y_1(\sqrt s L)}\left (1+O(e^{-2q_I z_1})\right )\cr
&=&\frac{H_2^{(1)}(\sqrt s L)}{H_1^{(1)}(\sqrt s L)}\left (1+O(e^{-2q_I z_1})\right )
\eea
where we employed the definition of the Hankel functions. Substituting this result into eq.~(\ref{totalprop}) we find, 
up to $O(e^{-2q_I z_1})$ terms
\beq
\langle h_{\mu \nu} h_{\rho\sigma}\rangle (\sqrt s= q_R+iq_I)=-\frac{H_2^{(1)}(\sqrt s L)}{2\sqrt s H_1^{(1)}(\sqrt s L)}
\Pi_{\mu\nu\rho\sigma}^{m\not = 0} 
-\frac{\eta_{\mu\nu}\eta_{\rho\sigma}}{ 6 s L}.
\label{euclideanampl}
\eeq
The important point about this result  is that all the dependence on powers of $z_1$ has disappeared. 
Now, the point is that if we use initial
states that have an energy spread $(\Delta s)/\sqrt s>> 1/ z_1$ (i.e. smearing this amplitude with a  wave function spread 
over $\Delta s$)
is practically equivalent to considering the amplitude at a complex point with ${\rm Im} (\sqrt s)\sim i \Delta s/ \sqrt s$.
For such spread states all the physics of the KK-modes, including their detailed strong dynamics is exponentially suppressed, and thus
practically inaccessible. This result is quite analogous to the well known example of $e^+e^-\to hadrons$.
In that case, the behaviour of the cross
section as a function of the energy can be very complicated by the presence of the resonances, indicating that a perturbative
 QCD computation  is not adequate.  For instance this is the case near the bottom quark threshold. 
However by averaging the cross section over $\Delta s/ \sqrt s>> \Lambda_{QCD}$ one obtains an observable which can
be reliably computed in perturbation theory in terms of the production of quarks and gluons. This is the so-called parton-hadron
duality.

To conclude we want to apply eq.~(\ref{euclideanampl}) in the fully euclidean region $\sqrt s= iq$ to compute the gravitational
potential induced on the Planck brane by a source on it. Notice that this is the exact propagator for RSII, while for RSI
it applies only for $q> 1/z_1$.
We are interested in the long distance behaviour 
$q L\ll 1$, so we use the Bessel function expansion at small argument. 

We find
\beq
\langle h_{\mu \nu} h_{\rho\sigma}\rangle (\sqrt s=+iq)=\frac{1}{ M_5^3 L}\left\{\frac{\Pi_{\mu\nu\rho\sigma}^{m = 0}}{q^2}+
 \Pi_{\mu\nu\rho\sigma}^{m\not = 0} \frac{L^2}{2} \ln q^2 +O(q^2)\right \}
\label{forceonplackF}
\eeq
so that the Newtonian potential induced by a body of mass $m$ is
\beq
V(r)=\frac{G_N m}{r}\left ( 1 + \frac{2 L^2}{3 r^2}+\dots\right ).
\label{forceonplanck}
\eeq
This formula applies in the long distance regime $r\gg L$. 
 A good exercise is to check that at a distance $r\ll L$, the potential goes back to the 5D behaviour $V\sim 1/r^2$.
Then RSII can be a viable alternative to compactification for $L\lsim 100 \mu {\rm m}$.

\section*{An afterword}

These lectures were delivered at Carg\`ese in the summer of 2003  to a mixed audience of theoretical and experimental Ph.D. students.
The aim had been  to present a pedagogical introduction to the subject
of extra dimensions, which, while being  theoretical in the approach, 
could be of value to both theorists and experimentalists. I hope I succeeded in that! In trying to be pedagogical 
I was obliged to select a  few topics and leave out some; for instance, I could not discuss the
 applications of extra dimensions to such directions of investigation  as cosmology, flavor physics, gauge-coupling unification, electroweak symmetry breaking. However, I believe I covered the
basic conceptual tools that are needed to understand on-going research on extra dimensions, with the  single important exception of holography.
This concerns the remarkable possibility of interpreting any 
5-dimensional model compactified on a slice of AdS5 as some purely
4-dimensional strongly coupled theory, where gravity is only ``weakly gauged''. Holography was originally born within string theory \cite{maldacena,witten} and its relevance to the understanding of
warped compactifications like the RS model was later discussed
in a number of more phenomenological papers \cite{maldacena2,apr,Rattazzi:2000hs,Perez-Victoria:2001pa}.
The holographic viewpoint is by now an essential part of the model-builder toolkit, and is, for instance,  central to the full understanding of some interesting recent developments of the RS model, where the electroweak symmetry-breaking dynamics is to a good extent calculable \cite{Agashe:2004rs}. 
Section \ref{strong} was written as a preparation to the discussion on holography. Unfortunately I could never find the time to write this discussion, and by now (summer 2006)  the phenomenological applications of holography have grown so much  that
it is worth devoting  a entire course  to them. Luckily this is already done in ref.  \cite{Gherghetta:2006ha}; those lectures,
together with refs. \cite{ apr,Rattazzi:2000hs}, and  with the more
``stringy'' lectures on AdS/CFT \cite{Maldacena:2003nj} are a good way to learn the applications of holography to phenomenology and model building.

\newpage

\def\ijmp#1#2#3{{\it Int. Jour. Mod. Phys. }{\bf #1~}(19#2)~#3}
\def\pl#1#2#3{{\it Phys. Lett. }{\bf B#1~}(19#2)~#3}
\def\zp#1#2#3{{\it Z. Phys. }{\bf C#1~}(19#2)~#3}
\def\prl#1#2#3{{\it Phys. Rev. Lett. }{\bf #1~}(19#2)~#3}
\def\rmp#1#2#3{{\it Rev. Mod. Phys. }{\bf #1~}(19#2)~#3}
\def\prep#1#2#3{{\it Phys. Rep. }{\bf #1~}(19#2)~#3}
\def\pr#1#2#3{{\it Phys. Rev. }{\bf D#1~}(19#2)~#3}
\def\np#1#2#3{{\it Nucl. Phys. }{\bf B#1~}(19#2)~#3}
\def\mpl#1#2#3{{\it Mod. Phys. Lett. }{\bf #1~}(19#2)~#3}
\def\arnps#1#2#3{{\it Annu. Rev. Nucl. Part. Sci. }{\bf #1~}(19#2)~#3}
\def\sjnp#1#2#3{{\it Sov. J. Nucl. Phys. }{\bf #1~}(19#2)~#3}
\def\jetp#1#2#3{{\it JETP Lett. }{\bf #1~}(19#2)~#3}
\def\app#1#2#3{{\it Acta Phys. Polon. }{\bf #1~}(19#2)~#3}
\def\rnc#1#2#3{{\it Riv. Nuovo Cim. }{\bf #1~}(19#2)~#3}
\def\ap#1#2#3{{\it Ann. Phys. }{\bf #1~}(19#2)~#3}
\def\ptp#1#2#3{{\it Prog. Theor. Phys. }{\bf #1~}(19#2)~#3}

\end{document}